\documentclass[letterpaper]{article} 
\usepackage{aaai25}  
\usepackage{times}  
\usepackage{helvet}  
\usepackage{courier}  
\usepackage[hyphens]{url}  
\usepackage{graphicx} 
\usepackage{natbib}  
\usepackage{caption} 
\usepackage{algorithm}
\usepackage{algorithmic}
\usepackage{amsmath}
\usepackage{amsfonts}
\usepackage{color}
\usepackage{subfigure}
\usepackage{marvosym}
\usepackage[urlcolor=blue]{hyperref}
\usepackage{newfloat}
\usepackage{listings}
\DeclareCaptionStyle{ruled}{labelfont=normalfont,labelsep=colon,strut=off} 
\floatstyle{ruled}
\newfloat{listing}{tb}{lst}{}
\floatname{listing}{Listing}
\title{\textcolor{red}{Dynamic PDB}: A New Dataset and a SE(3) Model Extension by \\ Integrating Dynamic Behaviors and Physical Properties in Protein Structures}
\author{
    Ce Liu\textsuperscript{\rm 1}\equalcontrib, 
    Jun Wang\textsuperscript{\rm 1}\equalcontrib, 
    Zhiqiang Cai\textsuperscript{\rm 1}\equalcontrib, 
    Yingxu Wang\textsuperscript{\rm 1,3}, 
    Huizhen Kuang\textsuperscript{\rm 2}, 
    Kaihui Cheng\textsuperscript{\rm 2},\\
    Liwei Zhang\textsuperscript{\rm 1},
    Qingkun Su\textsuperscript{\rm 1}, 
    Yining Tang\textsuperscript{\rm 2}, 
    Fenglei Cao\textsuperscript{\rm 1},
    Limei Han\textsuperscript{\rm 2},\\
    Siyu Zhu\textsuperscript{\rm 2 \Letter}, 
    Yuan Qi\textsuperscript{\rm 2 \Letter}
}
\affiliations{
    \textsuperscript{\rm 1}Shanghai Academy of Artificial Intelligence for Science\\
    \textsuperscript{\rm 2}Fudan University\\
    \textsuperscript{\rm 3}Mohamed bin Zayed University of Artificial Intelligence
}

\usepackage{bibentry}

\begin{document}
\maketitle

\begin{abstract}
Despite significant progress in static protein structure collection and prediction, the dynamic behavior of proteins, one of their most vital characteristics, has been largely overlooked in prior research. 
This oversight can be attributed to the limited availability, diversity, and heterogeneity of dynamic protein datasets.
To address this gap, we propose to enhance existing prestigious static 3D protein structural databases, such as the Protein Data Bank (PDB), by integrating dynamic data and additional physical properties.
Specifically, we introduce a large-scale dataset, Dynamic PDB, encompassing approximately 12.6K proteins, each subjected to all-atom molecular dynamics (MD) simulations lasting 1 microsecond to capture conformational changes. 
Furthermore, we provide a comprehensive suite of physical properties, including atomic velocities and forces, potential and kinetic energies of proteins, and the temperature of the simulation environment, recorded at 1 picosecond intervals throughout the simulations.
For benchmarking purposes, we evaluate state-of-the-art methods on the proposed dataset for the task of trajectory prediction. 
To demonstrate the value of integrating richer physical properties in the study of protein dynamics and related model design, we base our approach on the SE(3) diffusion model and incorporate these physical properties into the trajectory prediction process. 
Preliminary results indicate that this straightforward extension of the SE(3) model yields improved accuracy, as measured by MAE and RMSD, when the proposed physical properties are taken into consideration. 
\url{https://fudan-generative-vision.github.io/dynamicPDB/}
\end{abstract}

%

\section{Introduction}
\begin{figure*}[t]
\centering
\includegraphics[width=0.90\textwidth]{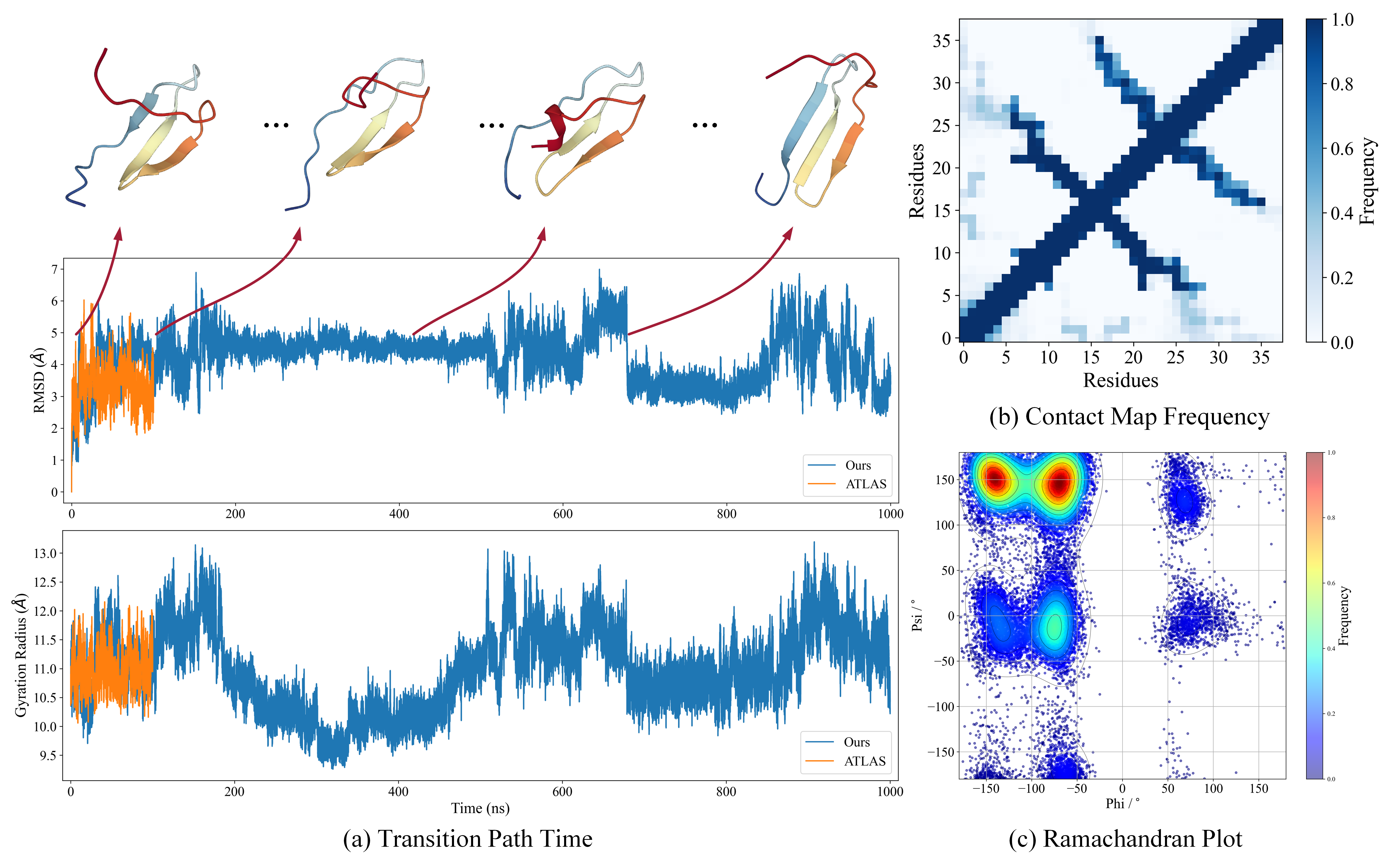}
\caption{The conformational evolution and statistics of protein 3TVJ\_I from proposed dataset. a) The regions with the most significant changes in the RMSD (Root Mean Square Deviation) and radius of gyration curves over time correspond to potential conformational changes, as depicted in the upper part of the figure. b) The contact map frequency illustrates the changes in interactions between residues within the protein. c) The Ramachandran plot provides insight into the dihedral angles of the protein backbone, indicating the structural validity of the protein conformation.}
\vspace{-4 mm}
\label{fig:conf_evo}
\end{figure*}
Advancements in experimental techniques such as X-ray crystallography, nuclear magnetic resonance, and electron microscopy have substantially enriched the repository of static 3D structural data for biological macromolecules, including proteins, nucleic acids, and complex assemblies~\cite{berman2000pdb}. 
This extensive repository not only facilitates research in the prediction of static protein structures but also plays a pivotal role in fundamental scientific inquiry, applied sciences, and technological development. 
Specifically, a comprehensive understanding of protein 3D structures is crucial for the rational design of drug molecules that precisely target specific active sites, thereby enhancing drug efficacy and minimizing adverse side effects~\cite{ zhang2009protein,sledz2018protein}. 
Moreover, the capacity to predict static protein structures can facilitate the development of novel enzymes, thereby improving the efficiency of biocatalytic reactions~\cite{dobson2005predicting}. 
Additionally, these structures serve as critical references for the synthesis of new functional proteins or biological modules, which in turn propels the advancement of innovative biological systems and products~\cite{pantazes2011recent,huang2016coming,kuhlman2019advances}.

Among the various repositories of static protein structure data, the Protein Data Bank (PDB)~\cite{berman2000pdb} is particularly noteworthy, encompassing over 220K proteins with experimentally determined 3D structures acquired through methodologies such as X-ray crystallography and cryo-electron microscopy. 
The advent of deep learning techniques, as exemplified by AlphaFold~\cite{jumper2021highly} and its derivatives~\cite{baek21rose, wu2022omega,rives2019biological,zeming23llm}, has revolutionized the domain by accurately predicting the structures of over one million proteins based on their amino acid sequences. 
However, the analysis of proteins’ dynamic behaviors, critical for understanding conformational changes, remains a substantial challenge. 
This challenge is primarily due to the limited availability, diversity, and heterogeneity of dynamic protein datasets~\cite{vander23atlas, vander2021medusa, marchetti2021machine}.

This paper introduces a novel dataset, Dynamic PDB, designed to capture the dynamic behavior of proteins, accompanied by a comprehensive suite of physical properties such as atomic velocities and forces, potential and kinetic energies, and the temperature of the simulation environment. 
To address the challenges inherent in studying protein dynamics, molecular dynamics (MD) simulations have been extensively employed~\cite{lindahl2008membrane,klepeis2009long,lindorff2011fast}. 
These simulations monitor atomic motion within protein systems by integrating Newton's equations of motion over time. 
MD simulations have proven crucial in providing insights into protein conformational behavior at both local and global scales. 
A significant advantage of MD simulations is their ability to uncover allosteric pathways through simulations lasting hundreds of nanoseconds~\cite{rivalta2012allosteric, rivalta2016allosteric,wurm2021molecular}, while extended simulations enable the observation of substantial conformational changes~\cite{klepeis2009long,ayaz2023structural}. 
Specifically, protein structure ensembles generated from MD trajectories enhance docking performance~\cite{sledz2018protein}, reveal pockets involved in protein-protein interactions~\cite{jubb2015flexibility}, and elucidate the flexibility patterns of residues associated with protein interface formation~\cite{Kokkinidis12enzymology, teilum2009functional}.
Prior to the present work, existing datasets such as MoDEL~\cite{meyer10model}, Dynameomics~\cite{kehl08dynameomics}, and ATLAS~\cite{vander23atlas} have primarily offered general molecular dynamics (MD) data for soluble proteins. 
Our contributions provide three key advancements: (1) finer-grained time sampling intervals of 1 picosecond, allowing for the capture of more detailed allosteric pathways; 
(2) extended time sampling durations of up to 1 microsecond per protein, which facilitate a more comprehensive understanding of significant conformational changes; 
and (3) an enriched array of physical properties captured during the molecular dynamics process, including atomic velocities and forces, potential and kinetic energies, and the temperature of the simulation environment.

In this study, we evaluate state-of-the-art methods for trajectory extrapolation using our proposed dataset. 
Our findings indicate that finer-grained time sampling intervals significantly enhance the resolution of allosteric pathways during the trajectory extrapolation process. 
Meanwhile, extended time sampling intervals facilitate a more comprehensive understanding of critical conformational transitions. 
An example is shown in Figure~\ref{fig:conf_evo}.
To demonstrate the advantages of incorporating comprehensive physical properties into the analysis of protein dynamics and model design, we develop our approach based on the SE(3) diffusion formulation~\cite{yim23se3}. 
Specifically, we propose to control the generation process based on conditions of amino acid sequence and physical properties.
This formulation is augmented by integrating these physical properties into the trajectory prediction process. 
Preliminary results suggest that this straightforward extension of the SE(3) diffusion model improves accuracy, as indicated by MAE and RMSD, when the proposed physical properties are systematically incorporated into the analysis.

\section{Related Work}
This section commences with a review of the existing literature pertaining to protein structural datasets, as well as the associated tasks of structural prediction and conformation sampling. 
Following this, we present an overview of protein datasets that integrate molecular dynamics, accompanied by a discussion of the tasks involved in protein trajectory prediction based on these datasets.
\begin{figure*}[t]
\centering
\includegraphics[width=0.96\linewidth]{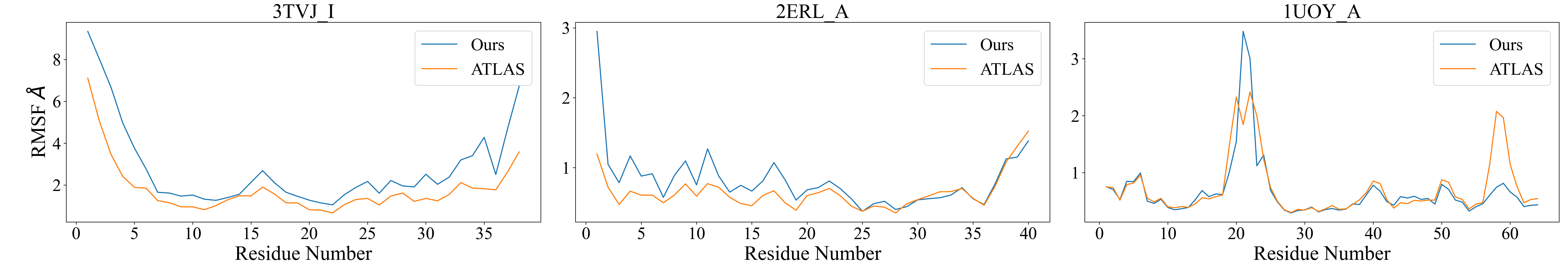}
\caption{RMSF comparison between the proposed dataset and ATLAS reveals similar residue fluctuations, effectively capturing the intrinsic dynamics of proteins.}
\label{fig:rmsf_comparison}
\end{figure*}


\paragraph{Protein Structural Dataset.}
The Protein Data Bank (PDB)~\cite{berman2000pdb} is a comprehensive repository of protein structures, currently encompassing 223.8K experimentally determined structures and 1.1M computational models from AlphaFoldDB~\cite{varadi21alphafolddb} and RoseTTAFold~\cite{baek21rose}. 
Additionally, the Structural Classification of Proteins (SCOP)~\cite{murzin1996structural} and CATH~\cite{sillitoe2021cath} databases provide classifications based on structural and evolutionary relationships. 
This extensive collection of 3D structures significantly influences various research fields, including \textit{de novo} protein design, protein structure prediction, and conformation sampling. 
In this study, we utilize the prestigous PDB dataset, integrating dynamic behaviors and physical properties.

The task of \textit{de novo} protein design~\cite{trippe2023diffusion} aims to generate novel protein structures with desired properties. 
For example, RFDiffusion~\cite{watson2023novo} generates protein backbones utilizing 
denoising diffusion probabilistic models (DDPM)~\cite{ho2020denoising, zhu2024champ, xu2024hallo}. 
Furthermore, $\mathrm{SE}(3)$-Diff~\cite{yim23se3} proposed $\mathrm{SE}(3)$ invariant diffusion models to learn $\mathrm{SE}(3)$ equivariant scores based on denoising score matching~\cite{song2020score}.

The prediction of protein structures from amino acid sequences is a formidable challenge in biology. 
A variety of approaches, including molecular simulations~\cite{brooks2002protein,snow2005well} and bioinformatics~\cite{roy2010tasser,marks2011protein}, have been employed to address this issue. 
Notably, AlphaFold~\cite{jumper2021highly} represents a significant advancement, leveraging innovative attention mechanisms and training protocols to markedly improve predictive accuracy. 
RoseTTAFold~\cite{baek21rose} enhances these concepts with a three-track network architecture, achieving further advancements in structure prediction accuracy.

Additionally, AlphaFlow~\cite{jing2023alphaflow}, Str2Str~\cite{lu2024strstr}, and ConfDiff~\cite{wang24cv} aim to sample diverse protein conformations by utilizing protein generation models.
\begin{figure*}[t]
\centering
\includegraphics[width=0.96\linewidth]{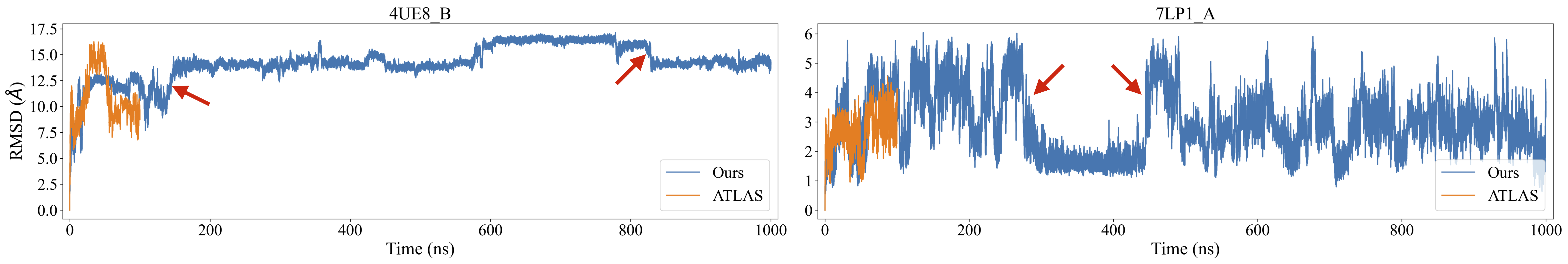}
\caption{RMSD plots for our dataset and ATLAS. Longer simulation time can potentially capture more protein conformational changes, which are indicated by the red arrows.}
\vspace{-4 mm}
\label{fig:rmsd}
\end{figure*}

\paragraph{Protein Dataset with Molecular Dynamics.}
Several public datasets have been established to study the dynamics of proteins. 
Notable examples include MemProtMD~\cite{newport2019memprotmd}, which focuses on membrane proteins; GPCRmd~\cite{rodriguez2020gpcrmd}, centered on G protein-coupled receptors (GPCRs); and SCoV2-MD~\cite{torrens2022scov2}, which is dedicated exclusively to SARS-CoV-2 proteins.
However, these datasets exhibit limitations in terms of scope and coverage.
Other datasets provide information on molecular dynamics simulations of general proteins. 
For instance, MoDEL~\cite{meyer10model} offers 1,875 trajectories for 1,595 representative proteins under near-physiological conditions, with simulation times ranging from 1 ns to 1 \textmu s; however, the platform is only partially functional and has not been updated recently. 
Dynameomics~\cite{kehl08dynameomics} examines the native and high-temperature dynamics of over 2,000 proteins and peptides in aqueous environments, yet it is currently inaccessible. 
ATLAS~\cite{vander23atlas} contains MD trajectories for 1,390 protein chains in solution, with simulations lasting 100 ns, but lacks comprehensive details such as atom-wise velocity and force. 
In contrast, our proposed dataset addresses these deficiencies by encompassing a wide variety of proteins from the Protein Data Bank 
and providing comprehensive information regarding protein dynamics, including velocity, force, energy, temperature, and other critical factors that influence a molecular dynamics system.

\paragraph{Protein Trajectory Prediction.}
Recently, coarse-grained (CG) molecular dynamics simulation has gained popularity~\cite{joshi2021review}. 
Flow Matching (FM) leverages normalizing flow techniques to generate samples and forces derived from the CG equilibrium distribution, facilitating the training of CG force fields~\cite{kohler2023flow}. 
Additionally, Denoising Force Field (DFF) employs score-based generative models to approximate CG force fields~\cite{arts2023two}. Nevertheless, CG methods fall short in providing detailed dynamical behaviors at the atomic level.
To approximate MD simulations for general molecules, several advanced network architectures have been proposed. 
The SE(3)-Trans~\cite{fuchs2020se} and Equivariant Graph Neural Networks (EGNN)~\cite{satorras2021n} have been specifically designed to exhibit equivariance under rotations and translations of point clouds. 
The Steerable E(3) Equivariant Graph Neural Network (SEGNN)~\cite{brandstetter2021geometric} further enhances these capabilities by incorporating steerable attributes and multi-layer perceptrons to account for geometric and physical quantities. 
The Second-order Equivariant Graph Neural Ordinary Differential Equation (SEGNO)~\cite{liu24segno} applies second-order motion laws to learn continuous trajectories. 
Furthermore, DiffMD~\cite{fang2023diffmd} introduces an equivariant geometric Transformer to learn the score function within the diffusion process. 
Given the proposed dataset, we can extend the SE(3) formulation~\cite{yim23se3} to integrate dynamic behaviors and physical properties, thereby improving the predictive capabilities for protein trajectories compared to previous approaches. 
We also anticipate that the proposed dataset will facilitate additional research analyses concerning the dynamic behaviors of protein structures.
\section{Dynamic Protein Dataset}
In this section, we introduce the proposed dataset, detailing the preparation process, the molecular dynamics simulation, and the analysis of dynamic behaviors.


    


\subsection{Source of Protein Structural Data}
Currently, there are 223.8K experimentally determined protein structures deposited in the Protein Data Bank (PDB)~\cite{berman2000pdb}. 
Despite the extensive scope of the PDB database, some structures, such as membrane proteins, present challenges for molecular dynamics (MD) simulations~\cite{lindahl2008membrane}. 
To address these challenges, we implement a series of processing steps, detailed below, to prepare the proteins for MD simulation.

\subsection{Preprocessing of Protein Data}
To prepare proteins for molecular dynamics simulations, we adhere to a systematic approach involving selection, cleaning, and completion. The flowchart is in appendix. 

\paragraph{Selection.} 
Initially, we select proteins with structures determined via X-ray diffraction, ensuring a resolution of no greater than 2.0 \AA. 
Subsequently, we filter out single-chain proteins that are monomeric in their oligomeric state and possess a sequence length of 500 residues or fewer. 
We further exclude membrane proteins by consulting databases such as OPM~\cite{lomize2006opm}, PDBTM~\cite{kozma2012pdbtm}, MemProtMD~\cite{newport2019memprotmd}, and mpstruc~\cite{bittrich2022rcsb}. 
Lastly, we utilize the dictionary of secondary structure in proteins (DSSP)~\cite{kabsch1983dictionary} to eliminate proteins characterized by more than 50\% coil-loop conformations.

\paragraph{Cleaning.} 
In this step, we remove all heteroatoms from the protein structures, including water molecules and ligands, to focus on the intrinsic dynamic behavior of the proteins. 
Additionally, we standardize the nomenclature by replacing non-standard residue names with their corresponding standard designations.

\paragraph{Completion.} 
For proteins exhibiting 5 or fewer missing residues, we employ MODELLER~\cite{webb2016comparative} for reconstruction. 
Conversely, for proteins with 6 or more missing residues, restoration is achieved utilizing AlphaFold 2~\cite{jumper2021highly}. 
Furthermore, we incorporate hydrogen atoms into the protein structures using MODELLER to ensure completeness.



\begin{figure}[t]
\includegraphics[width=1.0\linewidth]{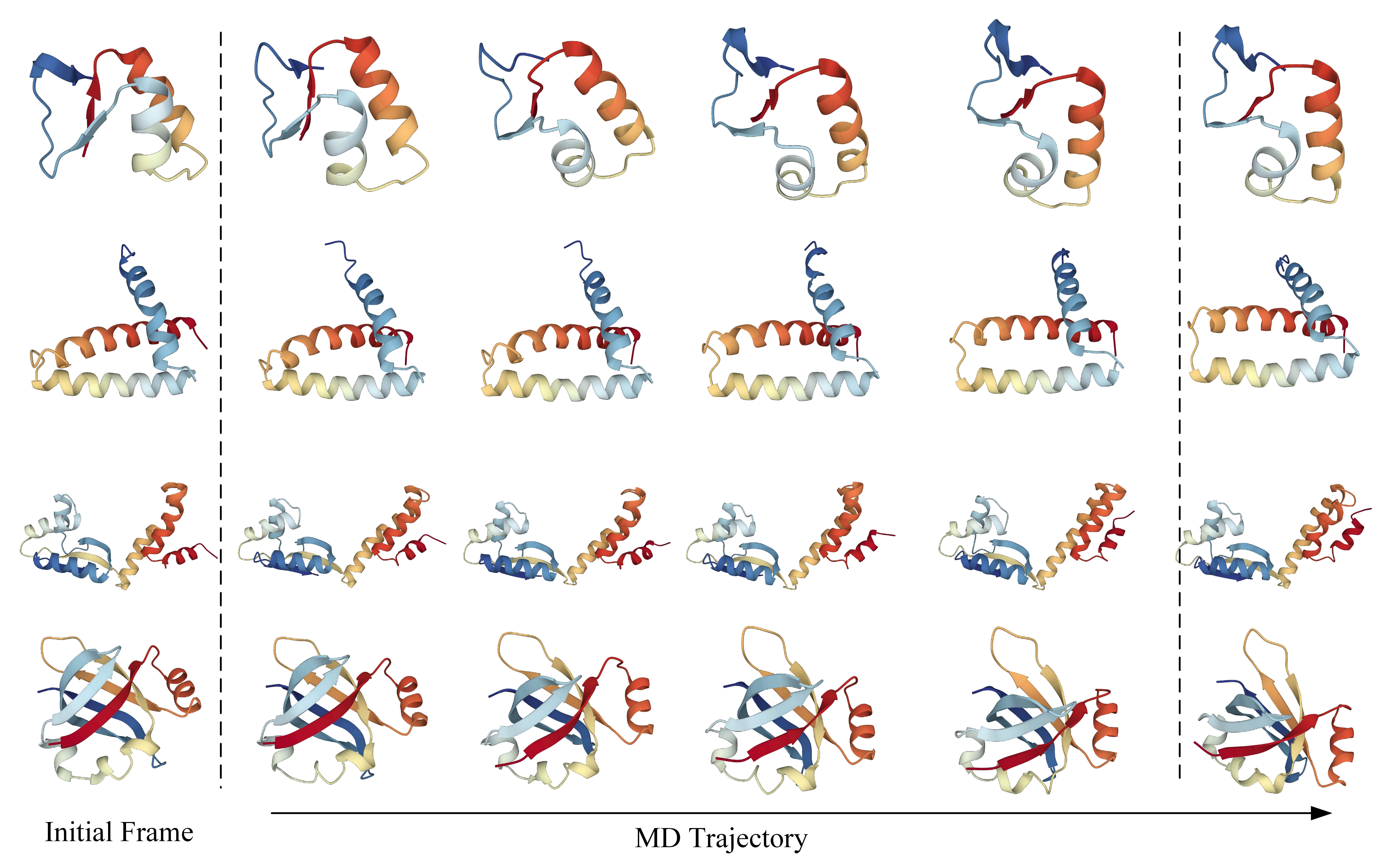}
\caption{Visualization of our protein trajectories, stored with higher temporal resolution, offers a more detailed depiction of the protein's trajectories.}
\vspace{-4 mm}
\label{fig:high_resolution}
\end{figure}
\subsection{MD Simulation Protocol}
All-atom molecular dynamics simulations are conducted using OpenMM~\cite{eastman17openmm} in conjunction with the Amber-ff14SB force field, which has been shown to enhance the accuracy of protein side chain and backbone parameters. 
The dimensions of the periodic box containing each protein are defined by a padding distance of 1 nm. 
This box is filled with TIP3P water molecules and subsequently neutralized with $\text{Na}^+/\text{Cl}^-$ ions at a concentration of 150 $mM$.

To mitigate structural artifacts, minimize atomic clashes, and establish a stable starting conformation for reliable simulation outcomes, an energy minimization process is performed. 
The force tolerance for this minimization is set to $2.39 \,kcal/mol \cdot nm$ without any imposed maximal step limits. 
Following energy minimization, two sequential equilibration processes are carried out: first in a canonical ensemble (NVT) and subsequently in an isothermal-isobaric ensemble (NPT), each spanning 1 ns with a time step of 1 fs. 
The \textit{LangevinMiddleIntegrator} is employed as the integrator for both equilibration phases, with the heat bath temperature and friction coefficient set to 300 $K$ and 1.0 ps$^{-1}$, respectively. 
During the NPT equilibration phase, the pressure is maintained at 1.0 bar utilizing the Monte Carlo Barostat.

Post-equilibration, the primary molecular dynamics simulations are executed, with each protein being simulated for a duration of 1 $\mu$s. 
A short time step of 1 fs is utilized to ensure computational stability throughout the simulation process. 
Atomic coordinates and various physical properties, including energy, velocity, and force, are recorded every 1 ps, yielding a total of 1.0M frames of data.

The simulations were performed on the GPU platform utilizing multiple processes, each running on a 16-core Intel Xeon CPU operating at 2.90 GHz, and supported by an NVIDIA A100 GPU with 80 GB of memory. 
The raw data generated from the MD trajectories and associated physical properties for the selected proteins form the foundation of our dataset.

\subsection{Analysis of Dynamic Protein Dataset}

\paragraph{Attributes of Proposed Dataset.}
The proposed dataset comprises extensive molecular dynamics simulation data for 12,643 distinct protein chains, meticulously organized to enhance accessibility and facilitate analysis. 
Statistics regarding protein length and secondary structures are presented in appendix, which illustrates the structural diversity present within the dataset. Specifically, protein lengths range from 9 to 444 residues, while the percentage of alpha helices varies from 0.00 to 0.92.

As an advantage of our dataset, the long-term simulations, illustrated in Figure~\ref{fig:conf_evo}, can potentially capture a broader range of protein conformational changes, whereas shorter sampling intervals, depicted in Figure~\ref{fig:high_resolution}, provide a more detailed representation of the protein’s trajectories.


Table~\ref{tab:simulationoutput} provides a detailed overview of the data attributes associated with each protein. This structured format supports subsequent analyses and interpretations of the intrinsic dynamic behaviors and properties of the proteins within the dataset. To facilitate further exploration of the dynamic behaviors of the proteins, we present additional relevant metrics in the following sections.


\begin{table}[t]
\centering
\begin{tabular}{c|c} 
\hline
    Description & Unit\\
\hline
    Identifier of Protein & -- \\
    
    Trajectory Coordinates & \AA \\
    
    Trajectory Velocities & \AA/ps \\
    
    Trajectory Forces & $\text{kcal/mol} \cdot \text{\AA}$ \\
    
    System Potential Energy & kJ/mole \\
    
    System Kinetic Energy & kJ/mole \\  
    
    System Total Energy & kJ/mole \\ 
    
    System Temperature & K \\
    
    System Volume Forces & $\text{nm}^3$ \\ 
    
    System Density & g/mL \\

    Status for Prolongation& -- \\
    
\hline
\end{tabular}
\caption{Attributes of proposed dataset.}
\vspace{-5 mm}
\label{tab:simulationoutput}
\end{table}

\paragraph{Root Mean Square Fluctuation.} 
The root mean square fluctuation (RMSF) represents a crucial metric in molecular dynamics, quantitatively assessing the standard deviation of atomic positions over the simulation trajectory. 
In this analysis, we randomly selected six proteins to compare their RMSF values with those obtained from the ATLAS dataset, with the results presented in Figure~\ref{fig:rmsf_comparison}. The plots indicate that both datasets exhibit analogous fluctuation patterns across the residue sequence, suggesting consistency in the dynamic behavior of the proteins.


\paragraph{Root Mean Square Deviation.} 
A significant advantage of the proposed dataset over previously established protein datasets in molecular dynamics is its extended simulation duration, which facilitates a more comprehensive exploration of conformational changes in proteins. 
In Figures~\ref{fig:conf_evo} and \ref{fig:rmsd}, we present the root mean square deviation (RMSD) curves for proteins from both datasets. 
The RMSD quantifies the deviation of the protein structure from its initial conformation. 
Extended simulations reveal markedly enhanced conformational dynamics, thereby providing richer biophysical insights into protein behavior.


\paragraph{Ramachandran Plot.} 
The Ramachandran plot serves as a tool to assess whether the dihedral angles $\phi$ and $\psi$ of amino acid residues in a protein structure fall within acceptable regions. 
Using protein 3TVJ\_I as a representative example, we visualize the Ramachandran plot in Figure~\ref{fig:conf_evo}. 
The plot demonstrates that the dihedral angles observed during our simulation are appropriately positioned within the permissible regions, indicating the structural validity of the protein.

\subsubsection{Contact Map Frequency.} 
The contact map functions as a representation of the spatial relationships between all possible pairs of amino acid residues, formatted as a binary two-dimensional matrix. For any two residues $i$ and $j$, the $(i,j)$ element of the matrix is assigned a value of 1 if the two residues are situated within a distance threshold of 7.5 \AA, and 0 otherwise. 
Using protein 3TVJ\_I as a case study, we illustrate the contact map frequency across the entire trajectory in Figure~\ref{fig:conf_evo}, normalizing the values to the range $[0, 1]$. 
Our analysis reveals that many elements fall between 0 and 1, indicating that the 3D structures of the protein undergo significant changes throughout the simulation.

\paragraph{Gyration Radius.} 
The gyration radius is defined as the distance from an axis of a body to a point within the body at which the moment of inertia is equivalent to that of the entire body. 
We present a plot depicting the gyration radius as a function of simulation time in Figure~\ref{fig:conf_evo}, illustrating the changes in this parameter throughout the duration of the simulation.



\section{Physics Conditioned Dynamic $\mathrm{SE}(3)$ Diffusion}

In this section, we begin with an overview of the $\mathrm{SE}(3)$ diffusion model~\cite{yim23se3}, employed for the generation of protein backbones from random noise. 
We then enhance the $\mathrm{SE}(3)$ diffusion model by integrating embeddings of amino acid sequences and relevant physical properties to improve the denoising process.
This augmented framework is subsequently applied to the task of protein trajectory prediction.

\begin{figure}
    \centering
    \includegraphics[width=1.0\linewidth]{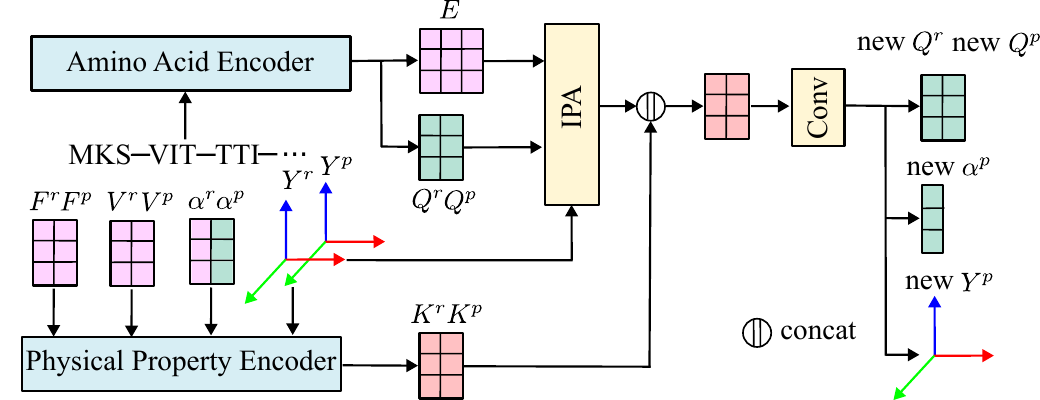}
    \caption{Overall architecture of our network. We first extract features by amino acid encoder and physical properties encoder respectively. Then we refine node features by IPA, and concatenate with the physical condition embedding. After 2D convolution operation, we predict the updated node features, torsion angles, and transformations.}
    \vspace{-4 mm}
    \label{fig:network_architecture}
\end{figure}

\subsection{$\mathrm{SE}(3)$ Diffusion Model}
The $\mathrm{SE}(3)$ model serves as a score-based generative approach~\cite{song2020score} for sampling protein backbones, characterized by a framework that parameterizes the backbone of a protein with $L$ residues.
This framework utilizes transformations $Y\in\mathrm{SE}(3)^{L}$ and torsion angles $\psi\in\mathbb{R}^{L}$.
The methodology incorporates a forward diffusion process on $\mathrm{SE}(3)^L$ to systematically transform the initial configurations $Y$ into a noise representation, followed by a reverse denoising process.
To achieve this, the $\mathrm{SE}(3)$ diffusion model employs a neural network to estimate both the score and the torsion angles. 
Initially, the model takes as input the node embedding, edge embedding, and a randomly initialized representation of $Y$.
A series of iterative layers is then applied to refine these embeddings and update $Y$. 
Notably, the final layer output predicts the torsion angles via the node embedding, while the transformation $Y$ is leveraged to compute the score.
The initialization of the node embedding is derived from residue indices and time steps, adhering to the methodology outlined in ProtDiff~\cite{trippe2023diffusion}, whereas the edge embedding incorporates relative sequence distances. 
Each layer of the model consists of an initial refinement of the node embedding using the Invariant Point Attention (IPA) module \cite{jumper2021highly}, followed by further processing with a spatial Transformer \cite{vaswani17attention}. 
Subsequently, the updated node embedding is utilized to enhance both the edge embedding and the transformation $Y$ via multi-layer perceptrons (MLPs).

\subsection{Integrating Dynamics and Physical Conditions}
In this section, we present an extension of the $\mathrm{SE}(3)$ diffusion model by integrating dynamic constraints and physical properties, with the overall architecture depicted in Figure~\ref{fig:network_architecture}. 
This enhancement of the $\mathrm{SE}(3)$ diffusion model involves the incorporation of the amino acid sequence alongside relevant physical characteristics to refine the denoising process. Additionally, we expand the representation of torsion angles to $\alpha\in\mathbb{R}^{L\times7}$, enabling the recovery of all atomic positions, including those of the side chains, in accordance with the methodologies established by AlphaFold 2.


\paragraph{Amino Acid Embedding.}
In our approach, we extract features from the amino acid sequence of the protein in accordance with the protein prediction framework established by OmegaFold~\cite{wu2022omega}.
For each residue $i$, we define a corresponding node feature $Q_{i}\in\mathbb{R}^{D_Q}$, which is designed to encapsulate critical information, such as the type of amino acid. 
Additionally, for each pair of residues $i$ and $j$, we introduce an edge feature $E_{(i,j)}\in\mathbb{R}^{D_E}$ that encodes the potential interactions between these residues. 
We enhance the original node and edge embeddings within the $\mathrm{SE}(3)$ diffusion framework by incorporating the features $Q$ and $E$.


\paragraph{Physical Condition Embedding.}
We propose a unified encoder to extract features from various physical properties. 
Specifically, for each residue $i$, we analyze the transformation $Y_{i}$, the torsion angles $\alpha_{i}$, the velocity $V_{i}\in\mathbb{R}^{3}$, and the force $F_{i}\in\mathbb{R}^3$ associated with the $\mathtt{C}_{\alpha}$ atom in residue $i$.
For each type of physical property, we initially apply MLPs to capture the relevant features. 
Subsequently, we normalize the extracted features across all residues using layer normalization~\cite{ba2016layer}. 
Finally, we concatenate the normalized features from all physical properties along the channel dimension. 
We denote the resultant feature vector that encapsulates all physical properties for residue $i$ as $K_{i}\in\mathbb{R}^{D_K}$.

\subsection{Network Overview} 
Finally, we apply our proposed design to the task of protein trajectory prediction. At the current time step, we denote the relevant variables, including transformation, torsion angles, velocity, and force, as $Y^r$, $\alpha^r$, $V^r$ and $F^r$ respectively. 
Correspondingly, we denote the variables at the next time step as $Y^p$, $\alpha^p$, $V^p$ and $F^p$.
The objective of this task is to estimate  $Y^p$ and $\alpha^p$ based on the known variables $Y^r$, $\alpha^r$, $V^r$ and $F^r$. 
To initiate the process, we first initialize the unknown variables at the next time step using the corresponding known variables from the current time step. 
For each residue and each time step, we extract both the amino acid embedding and the physical properties embedding. 
Subsequently, we refine the node features using the Invariant Point Attention (IPA) module and concatenate these features with the physical properties embedding. 
Finally, we apply 2D convolution along both the time dimension and the protein sequence dimension. 
The resulting node features at the next time step are then utilized to update the predicted transformation and torsion angles through multi-layer perceptrons (MLPs).

\section{Experiment}
In this section, we benchmark state-of-the-art methods on the proposed dataset. Additionally, we evaluate our proposed physics-conditioned dynamic SE(3) diffusion model, conducting ablation studies to demonstrate the efficacy of incorporating physical properties.

\subsection{Implementation}
\paragraph{Experimental Setup.}
All experiments were conducted on a server equipped with an Intel Xeon CPU operating at 2.90 GHz and an NVIDIA A100 GPU with 80 GB of memory. 
All methods were implemented using the PyTorch framework.

\paragraph{Training and Inference Details.}
Our method was trained with a batch size of 4, utilizing the AdamW optimizer for a total of 350,000 iterations. The learning rate was fixed at 1e-4. 
The weights assigned to the translation loss, rotation loss, and torsion loss were set to 100.0, 7.0, and 1.0, respectively. Additionally, we excluded other auxiliary losses originally proposed in the $\mathrm{SE}(3)$ diffusion framework. 
For the other methods, we employed their official released code along with the default hyperparameters.

\paragraph{Trajectory Prediction Task.}
We evaluate the methods in the context of the trajectory prediction task by selecting two specific proteins, 2ERL\_A and 3TVJ\_I. 
For each protein, we partition all time steps of the trajectory into training, validation, and test sets. 
Specifically, the first 70\% of the time steps are designated as the training set, the last 10\% serve as the test set, and the remaining 20\% are allocated for validation. 
Each method utilizes the available information at each time step as input to predict the 3D positions of atoms at the subsequent time step. 
For each method, we train a distinct set of parameters for each protein and report the average performance across both proteins.

\paragraph{Evaluation Metrics.}
We adopt two evaluation metrics to quantitatively compare the methods: the mean absolute error (MAE) and the root mean square deviation (RMSD).
For each protein, let the predicted coordinates be denoted as $S^{\mathtt{pred}}\in\mathbb{R}^{M\times 3}$, and the ground truth coordinates as $S^{\mathtt{gt}}\in\mathbb{R}^{M\times3}$, where $M$ represents the number of atoms. 
The coordinates are measured in angstroms \AA.  
To obtain a consolidated metric across multiple proteins, we weight the metric result for each protein by its respective number of residues. 
The MAE is calculated as follows: $\frac{1}{3M}\sum_i|S_i^{\mathtt{pred}}-S_i^{\mathtt{gt}}|$.
The RMSD is computed using the following expression: $\min_{R,T}\sqrt{\frac{1}{M}\sum_i|RS_i^{\mathtt{pred}}+T-S_i^{\mathtt{gt}}|^2}$.

\begin{table}[t]
\centering
\begin{tabular}{c|c|c} 
\hline
    Method & MAE & RMSD\\
\hline
    FM & 0.906 & 1.431\\
    SE(3)-Trans & 0.327 & 0.664 \\
    EGNN & 0.323 & 0.658\\
    SEGNN & 0.323 &  0.653 \\
    SEGNO & 3.902 & 9.780\\
\hline
    Ours & 0.290 & 0.352\\
\hline
\end{tabular}
\caption{Comparison to SOTA methpds for trajectory prediction on the proposed dataset.}
\vspace{-4 mm}
\label{tab:comparison_to_sota}
\end{table}

\subsection{Quantitative Results}
We compare our method with several state-of-the-art approaches, including FM~\cite{kohler2023flow}, SE(3)-Trans~\cite{fuchs2020se}, EGNN~\cite{satorras2021n}, SEGNN~\cite{brandstetter2021geometric}, and SEGNO~
\cite{liu24segno}, on the proposed dataset. The time interval between adjacent time steps is 1 ps.
The results of this comparison are presented in Table~\ref{tab:comparison_to_sota}.
It is important to note that FM predicts only the $\mathtt{C}_{\alpha}$ atoms; consequently, the metrics for FM are computed exclusively based on these atoms. 
Our method demonstrates superior performance across both evaluation metrics, achieving a significant reduction in RMSD from 0.653 to 0.352.
\\

\begin{figure}[t]
\subfigure[MAE]{
\begin{minipage}[b]{0.22\textwidth}
\includegraphics[width=1.0\textwidth]{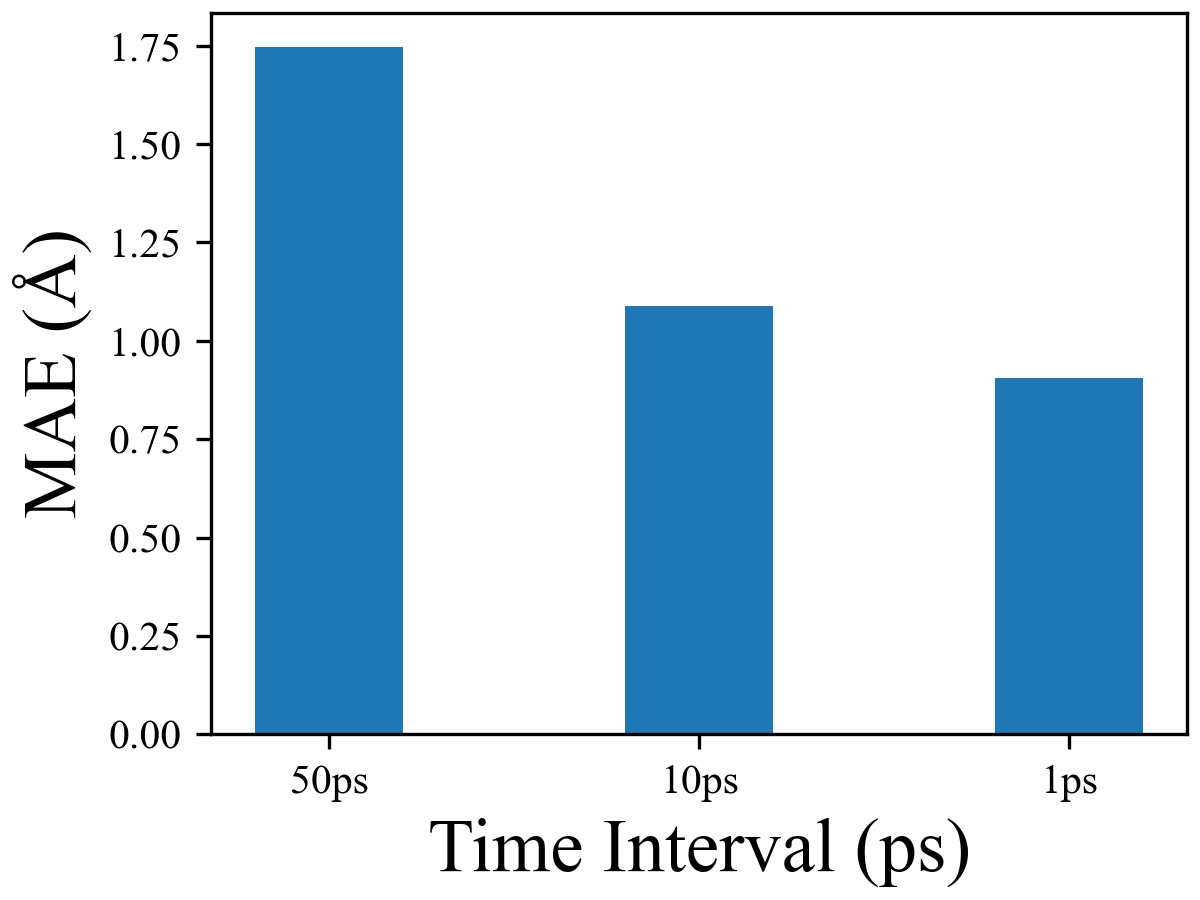}
\end{minipage}
}
\subfigure[RMSD]{
\begin{minipage}[b]{0.22\textwidth}
\includegraphics[width=1.0\textwidth]{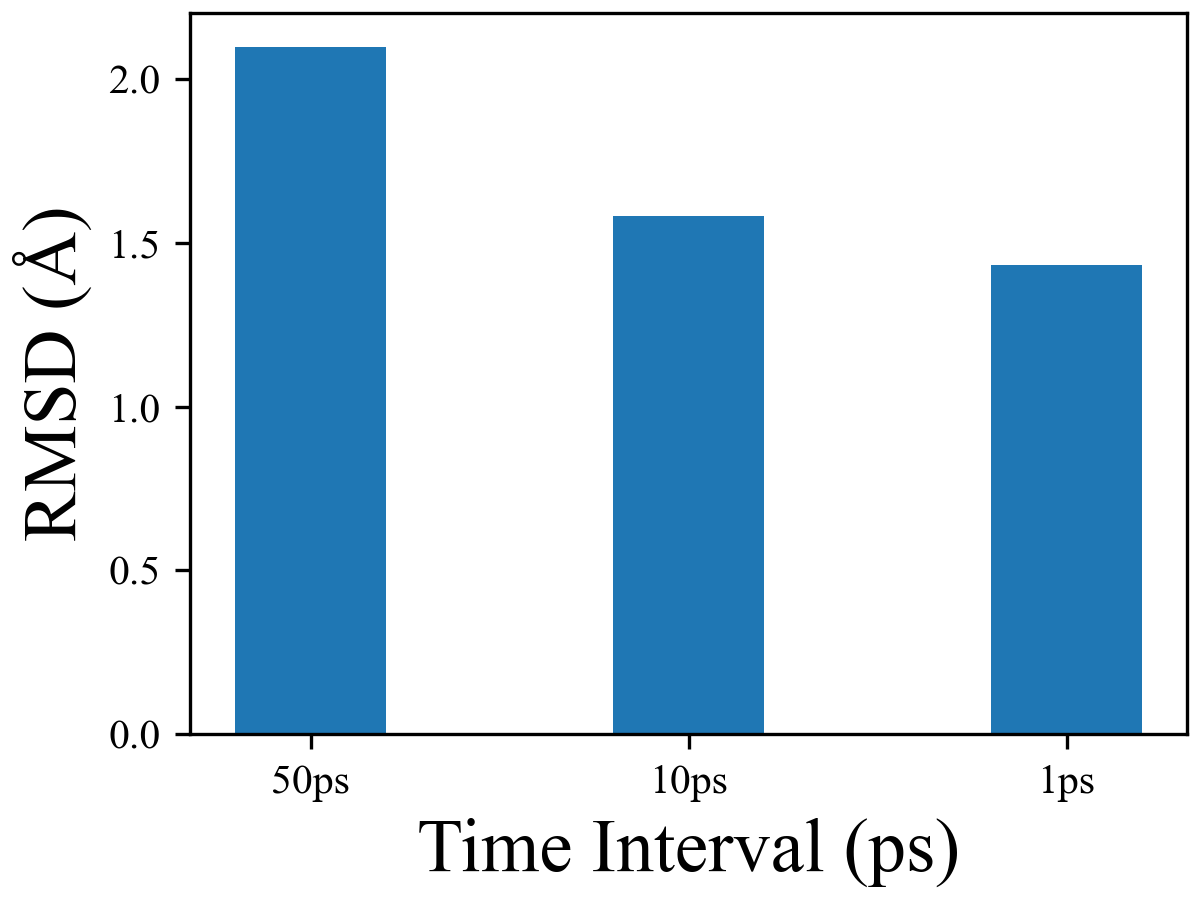}
\end{minipage}
}
\caption{Ablation study for the effects of time interval to the training of FM~\cite{kohler2023flow}. }
\vspace{-4 mm}
\label{fig:ablation_fm_time_interval}
\end{figure}

\paragraph{Analysis of Time Interval.}
The time interval between adjacent time steps significantly influences the prediction accuracy of the methods.
To investigate this effect, we evaluate the methods across different time intervals, specifically 1 ps, 10 ps, and 50 ps. 
The mean absolute error (MAE) results are reported in Table~\ref{tab:ablation_time_interval}.
Our findings indicate that all methods exhibit improved accuracy with shorter time intervals across both metrics. For instance, the MAE of EGNN increases by 97\% when the time interval is extended from 1 ps to 10 ps.
To further explore the impact of the time interval during training, we conducted an additional experiment using FM, which was trained on time intervals of 1 ps, 10 ps, and 50 ps, but evaluated exclusively at a 1 ps interval. 
The results of this evaluation are presented in Figure~\ref{fig:ablation_fm_time_interval}. Our analysis reveals that FM trained on shorter time intervals achieves superior accuracy.
\begin{table}[t]
\centering
\begin{tabular}{c|c|c|c} 
\hline
    Method & 1ps & 10ps & 50ps \\
\hline
    FM & 0.906 & 3.831  & 8.095\\
    SE(3)-Trans & 0.327  &  0.633 & 1.149\\
    EGNN & 0.323 & 0.639 & 1.181\\
    SEGNN & 0.323 & 0.621 & 1.133\\
    SEGNO & 3.902 & 3.958 & 4.111\\
\hline
\end{tabular}
\caption{Analysis of time interval on the proposed dataset.}
\vspace{-4 mm}
\label{tab:ablation_time_interval}
\end{table}

\begin{table}[t]
\centering
\begin{tabular}{c|c|c|c} 
\hline
    Method & 1us & 100ns & 50ns \\
\hline
    FM & 0.906 & 1.261 & 1.330\\
    SE(3)-Trans & 0.327 & 0.327 & 0.327\\
    EGNN &  0.323 &  0.327 & 0.327\\
    SEGNN &  0.323 & 0.323 & 0.321\\
    SEGNO &3.902 & 3.903 & 3.902\\
\hline
\end{tabular}
\caption{Analysis of simulation time on the proposed dataset.}
\label{tab:ablation_total_time}
\end{table}

\paragraph{Analysis of Simulation Time.}
In longer molecular dynamics (MD) simulations, a greater diversity of protein conformations is observed. 
To investigate this effect, we trained the methods on training sets generated from simulations lasting for 1 \textmu s, 100 ns, and 50 ns, respectively, while keeping the test set constant. The time interval between adjacent time steps is 1 ps.
The mean absolute error (MAE) results are reported in Table~\ref{tab:ablation_total_time}. 
Our findings indicate that FM exhibits improved accuracy with longer simulation durations. In contrast, SE(3)-Trans, EGNN, and SEGNN demonstrate comparable accuracy across different simulation lengths.

\paragraph{Analysis of Physical Properties.}
Physical properties, including velocity and force, provide valuable information for trajectory prediction. To investigate their effects, we conducted ablation studies using the protein 2ERL\_A. The time interval between adjacent time steps is 1 ps.
As shown in Table~\ref{tab:ablation_physics}, the inclusion of velocity resulted in a reduction of the mean absolute error (MAE) from 0.284 to 0.279 and the root mean square deviation (RMSD) from 0.546 to 0.533. 
Furthermore, by incorporating force into the model, both the MAE and RMSD were further decreased. 
These results underscore the beneficial impact of incorporating physical properties into the prediction process.

\begin{table}[t]
\centering
\begin{tabular}{c|c|c} 
\hline
    Method & MAE &  RMSD\\
\hline
    position & 0.284 & 0.546\\
    + velocity & 0.279 & 0.533\\
    + velocity and force & 0.277 & 0.528\\
\hline
\end{tabular}
\caption{Analysis of physical properties on proposed dataset.}
\vspace{-4 mm}
\label{tab:ablation_physics}
\end{table}

\subsubsection{Generalization on New Proteins} 
We train our diffusion model on trajectories of multiple proteins simultaneously, and make prediction for unseen proteins. The time interval between adjacent time steps is 1 ps.
Specifically, we train our model on 10 proteins and 214 proteins respectively, which exclude the protein 2ERL\_A, and then test on the 2ERL\_A as above, the results are shown in Table~\ref{tab:ablation_new_protein}. 
With more proteins during training, our model makes better prediction for new proteins.

\begin{table}[hbt]
\centering
\begin{tabular}{c|c|c} 
\hline
    \# Proteins & MAE & RMSD\\
\hline
    10 & 0.303 & 0.596\\
    214 & 0.299& 0.584\\
\hline
\end{tabular}
\caption{Results of the generalization test on new proteins.}
\vspace{-4 mm}
\label{tab:ablation_new_protein}
\end{table}

\subsection{Qualitative Results}
\begin{figure}[!h]
\centering
\subfigure[SE(3)-Trans]{
\begin{minipage}[b]{0.14\textwidth}
\includegraphics[width=1.0\textwidth]{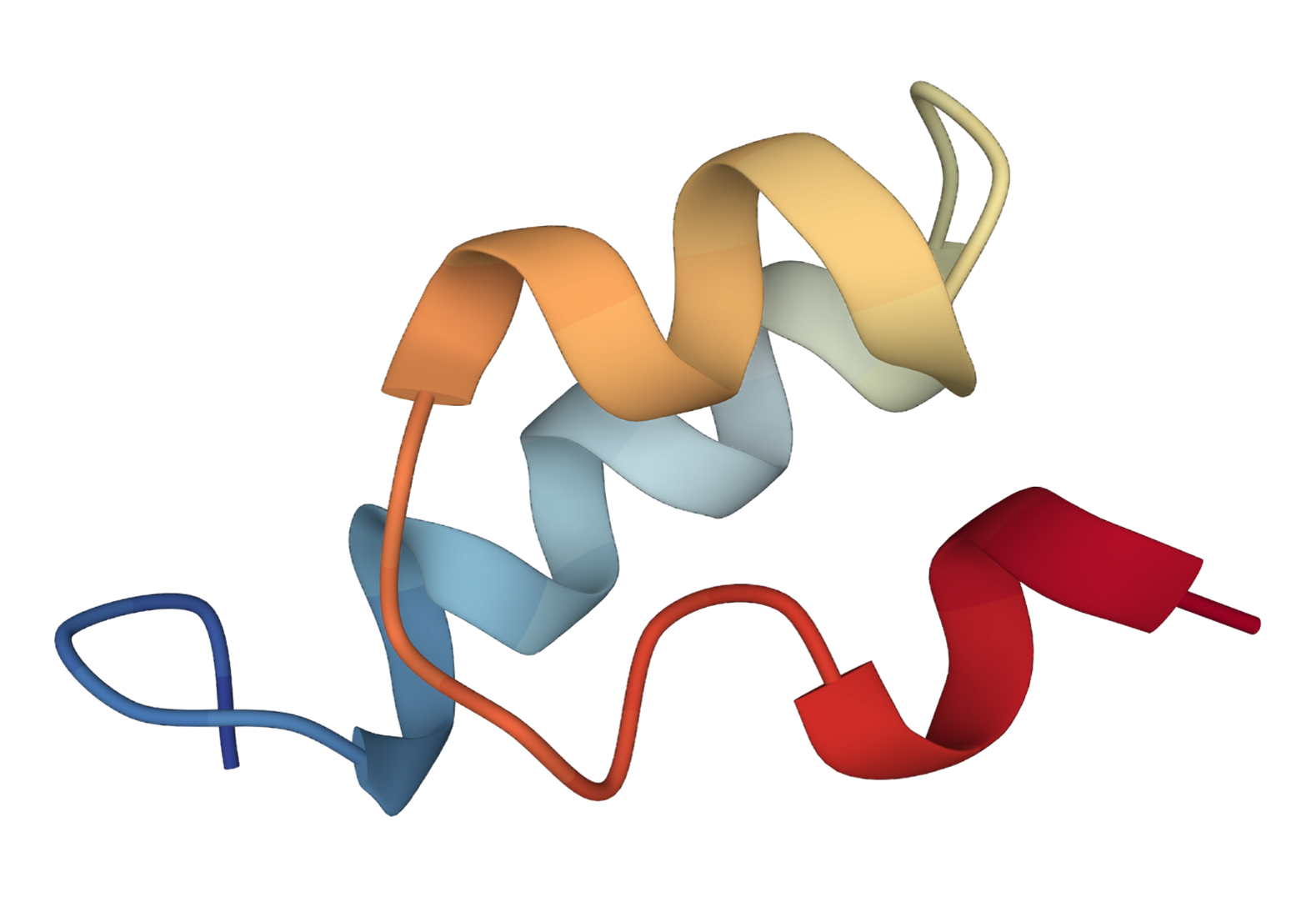}
\includegraphics[width=1.0\textwidth]{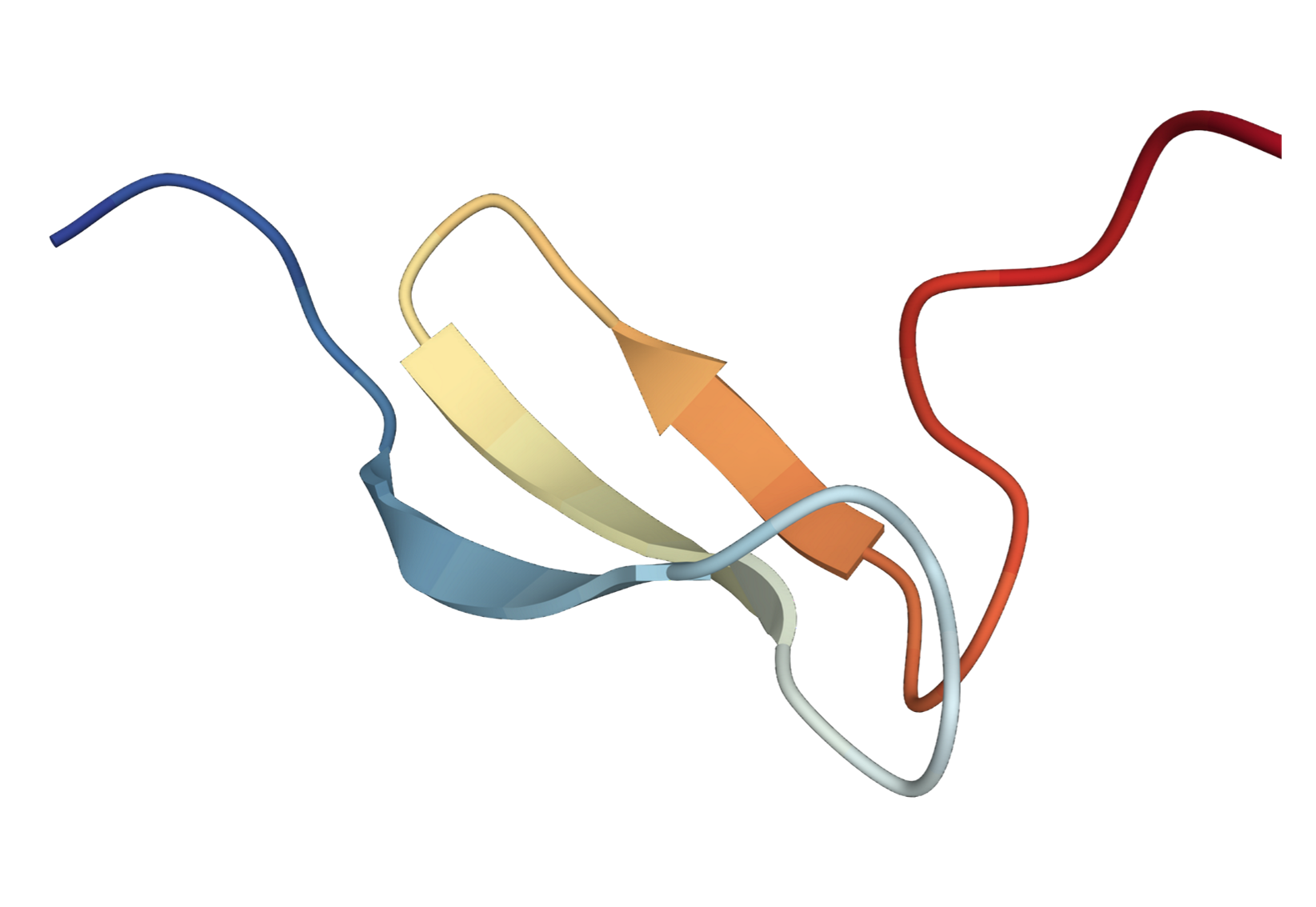}
\end{minipage}
}
\subfigure[Ours]{
\begin{minipage}[b]{0.14\textwidth}
\includegraphics[width=1.0\textwidth]{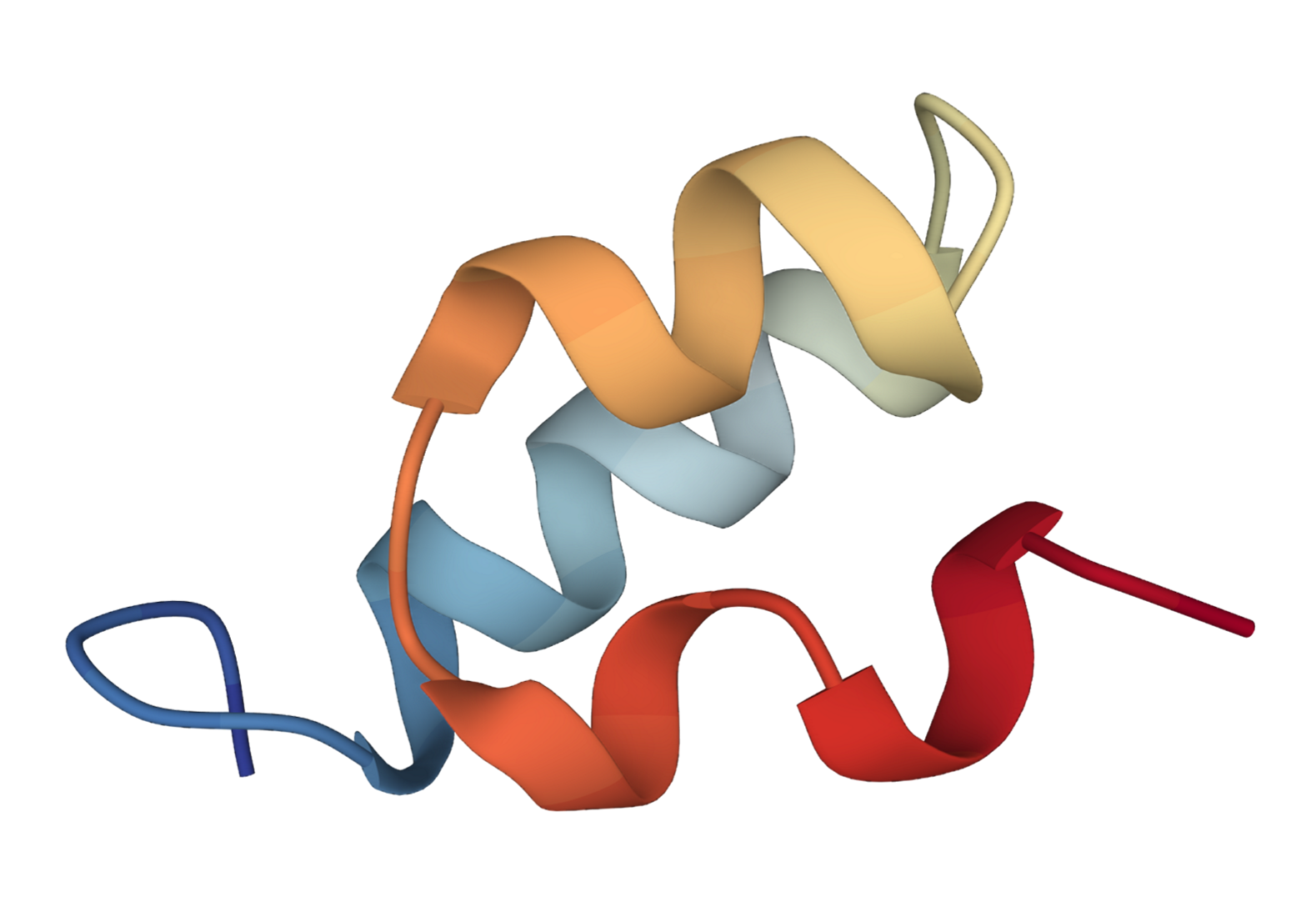}
\includegraphics[width=1.0\textwidth]{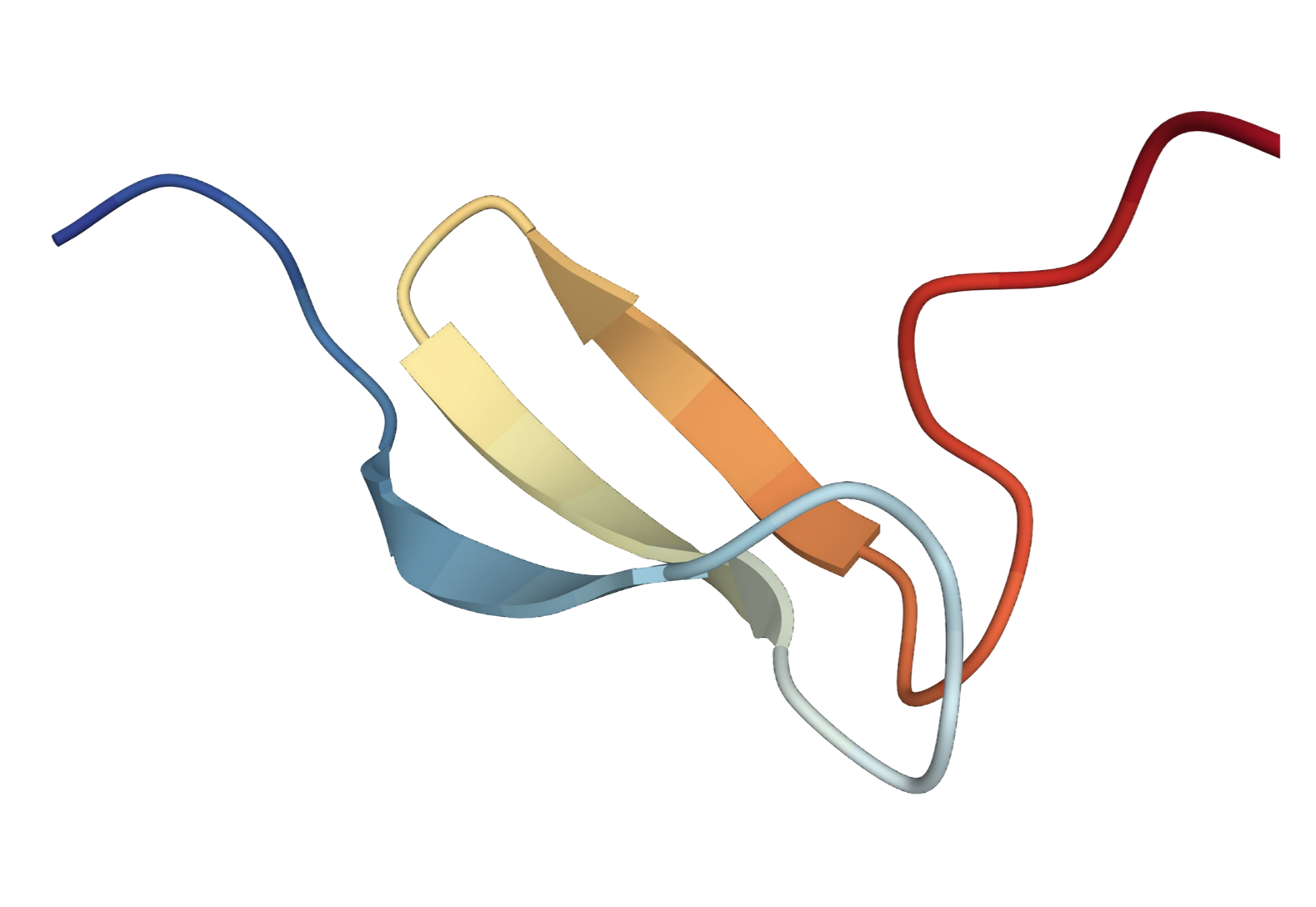}
\end{minipage}
}
\subfigure[GT]{
\begin{minipage}[b]{0.14\textwidth}
\includegraphics[width=1.0\textwidth]{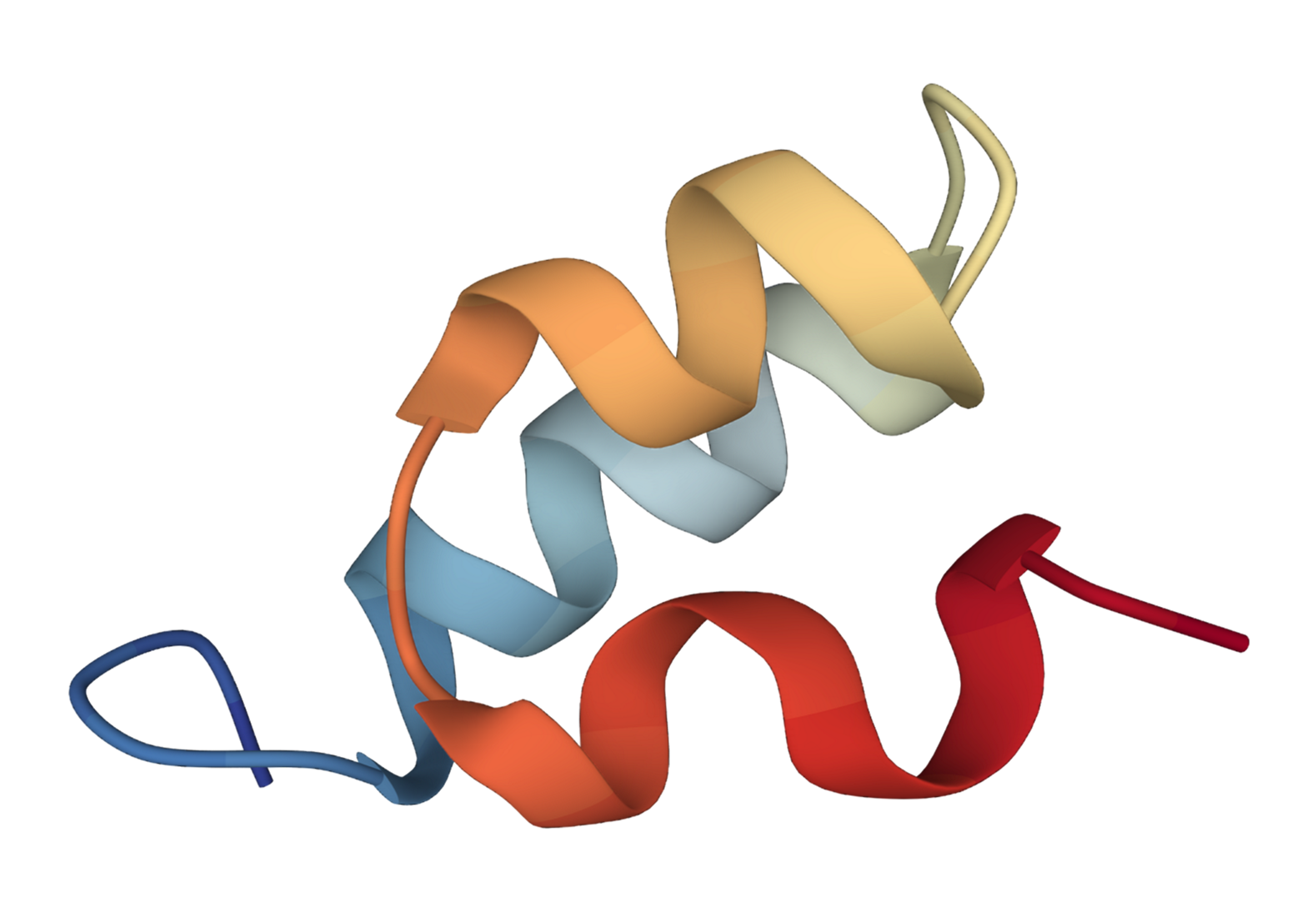}
\includegraphics[width=1.0\textwidth]{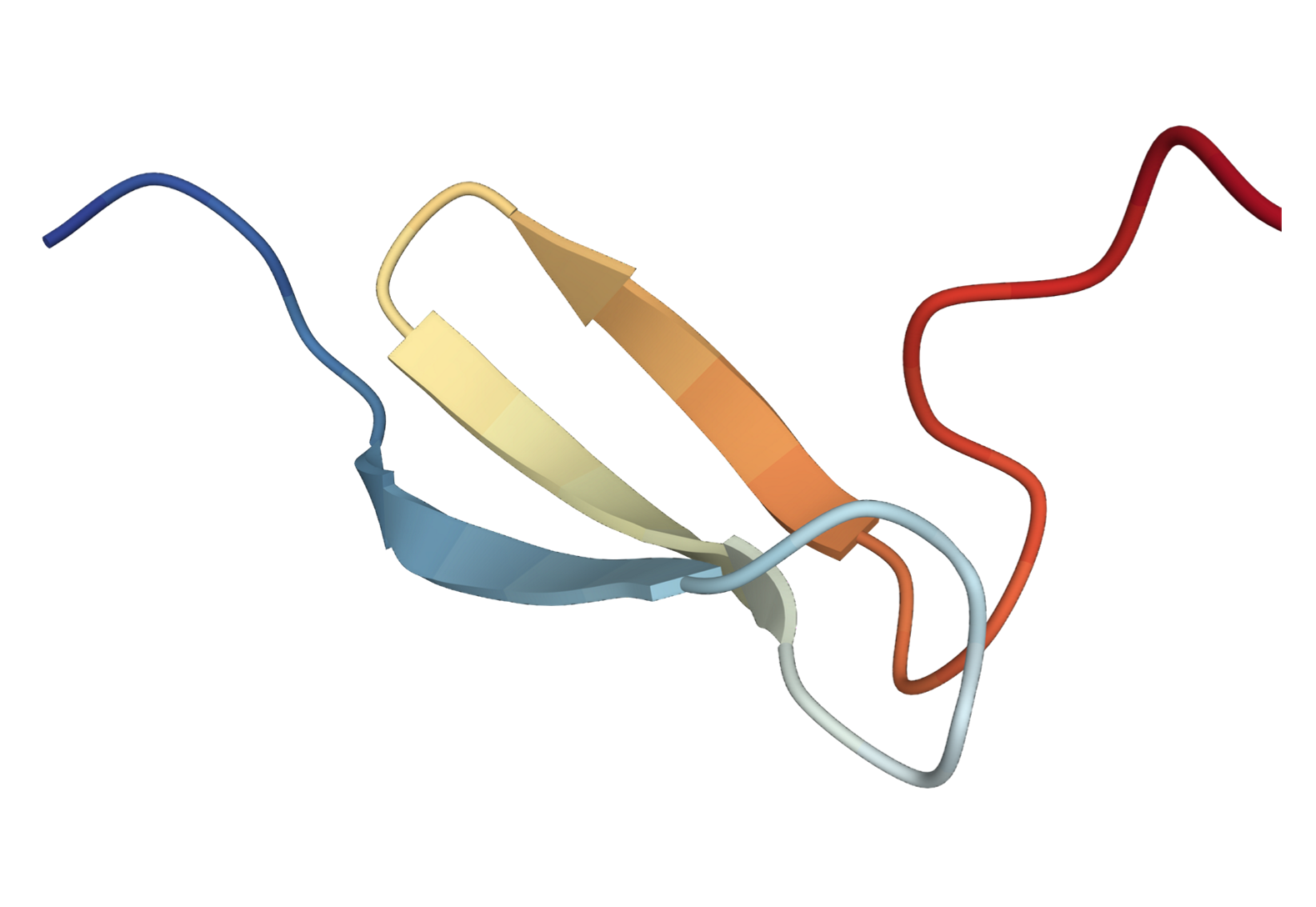}
\end{minipage}
}
\caption{Qualitative results on 2ERL\_A (top) and 3TVJ\_I (bottom). Our predictions are closer to the GT.}
\vspace{-4 mm}
\label{fig:qual_prediction}
\end{figure}

We visualize the predicted 3D structures from our method, and compare with SE(3)-Trans in Figure~\ref{fig:qual_prediction}. Our results demonstrate that the incorporation of physical properties allows the predictions from our approach to align more closely with the ground truth. The top two rows show the results on the protein 2ERL\_A, where our predictions on the alpha helices are more accurate. The bottom two rows show the predicted 3TVJ\_I. Our predictions are closer to the ground truth on the beta sheets.

\section{Limitations of Future Works}
While our dataset and model extension represent obvious advancements in capturing protein dynamics, several limitations warrant attention.
Firstly, although extensive, the dataset may not encompass the full diversity of protein conformations encountered in nature. 
Additionally, the computational expense associated with long-duration molecular dynamics simulations can be prohibitive, thus constraining the scalability of our approach.
Although the SE(3) model extension is effective, it may benefit from further refinement to fully leverage the integrated physical properties. 
Future research should prioritize expanding the dataset to include a broader array of proteins and conformational states. 
Moreover, optimizing computational efficiency and exploring alternative modeling techniques could enhance the predictive capabilities and overall applicability of our methodology.
Further investigations could also focus on integrating additional physical and biochemical properties, providing a more comprehensive understanding of protein dynamics.
\section{Conclusion}
In this paper, we presented a novel dataset that integrates dynamic behaviors and physical properties within protein structures, effectively addressing the limitations of existing static protein databases. 
By conducting all-atom molecular dynamics simulations on approximately 12.6K proteins, we captured significant conformational changes and compiled a comprehensive suite of physical properties.
Our extension of the SE(3) diffusion model, which incorporates these physical properties, demonstrated enhanced accuracy in trajectory prediction tasks, as indicated by reductions in mean absolute error (MAE) and root mean square deviation (RMSD). 
The results underscore the critical role of dynamic data and physical properties in advancing the understanding of protein dynamics.
This work paves the way for future research in protein trajectory prediction and the development of more sophisticated models that capitalize on dynamic protein datasets.

\nobibliography*
\appendix
\bibliography{aaai25}
\setcounter{figure}{0}
\renewcommand{\thefigure}{A.\arabic{figure}} 
\renewcommand{\thetable}{A.\arabic{table}}
\setcounter{table}{0}
\section{Appendix}

\subsection{More Details for Dynamic Protein Dataset}
\begin{figure}[h]
\centering
\includegraphics[width=0.65\linewidth]{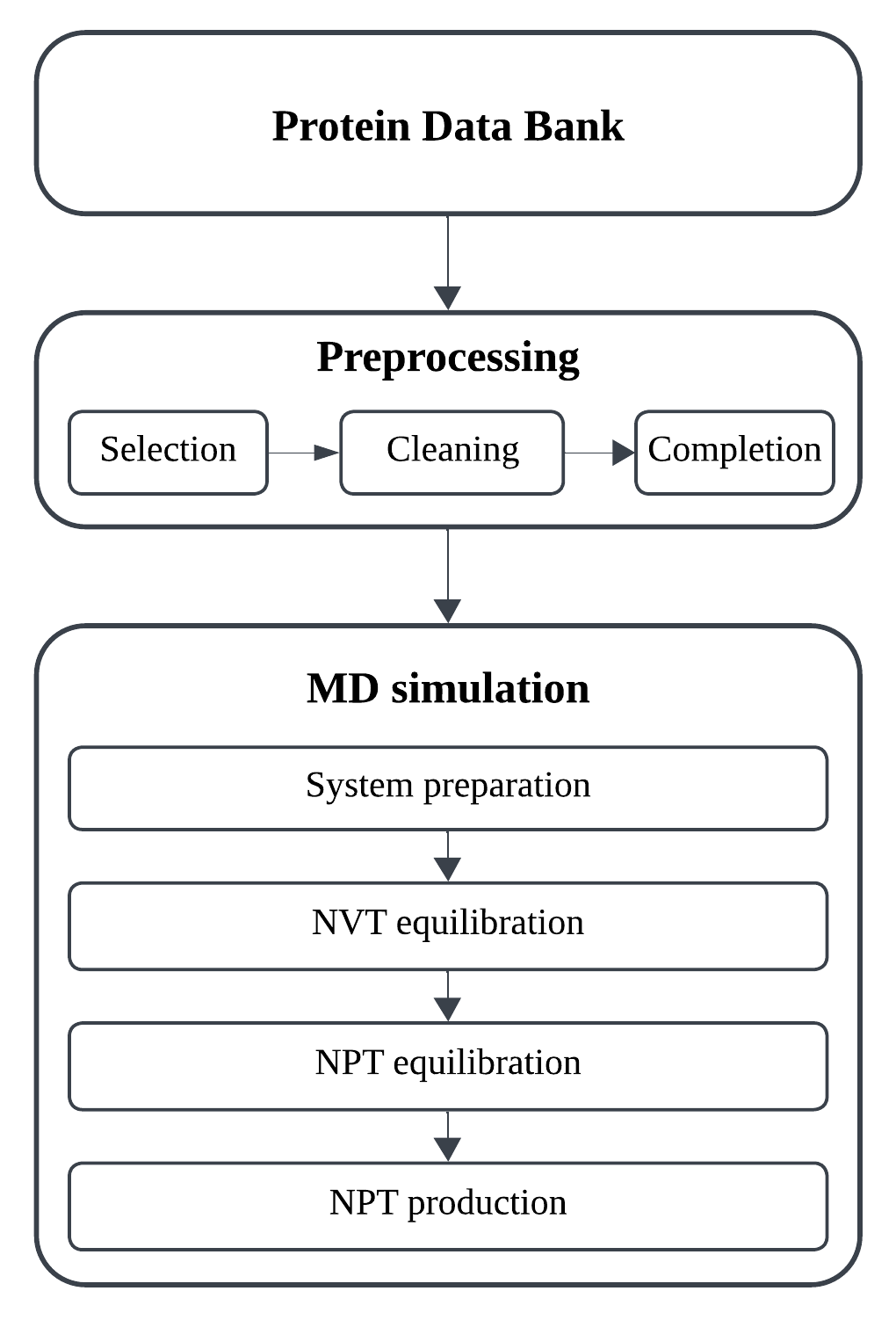}
    \caption{Flowchart of dynamic protein dataset construction.}
\label{fig:flowchart}
\end{figure}

\paragraph{Overall Pipeline.}
In this part, we present an overview of the pipeline in Figure~\ref{fig:flowchart}. The protein structure is from the Protein Data Bank. We first preprocess and then perform MD simulation. 

\begin{figure}[h]
\centering
\includegraphics[width=0.75\linewidth]{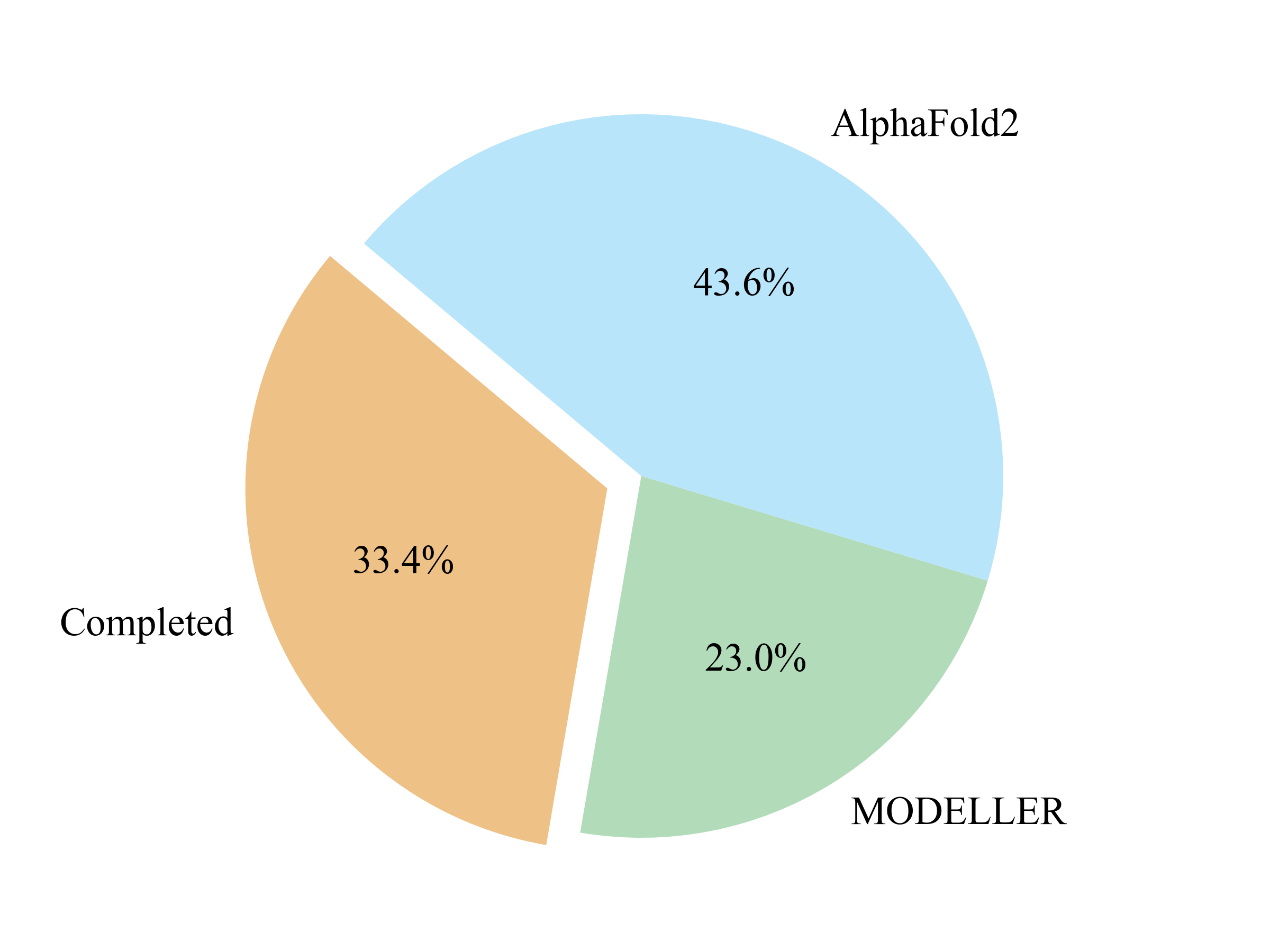}
    \caption{The percentage of proteins by different complementing methods. "Completed" means no missing residues.}
\label{fig:completion_percent}
\vspace{-1 mm}
\end{figure}

\begin{figure*}[t]
    \centering
    \begin{minipage}[b]{0.43\textwidth}
        \centering
        \includegraphics[width=1.0\textwidth]{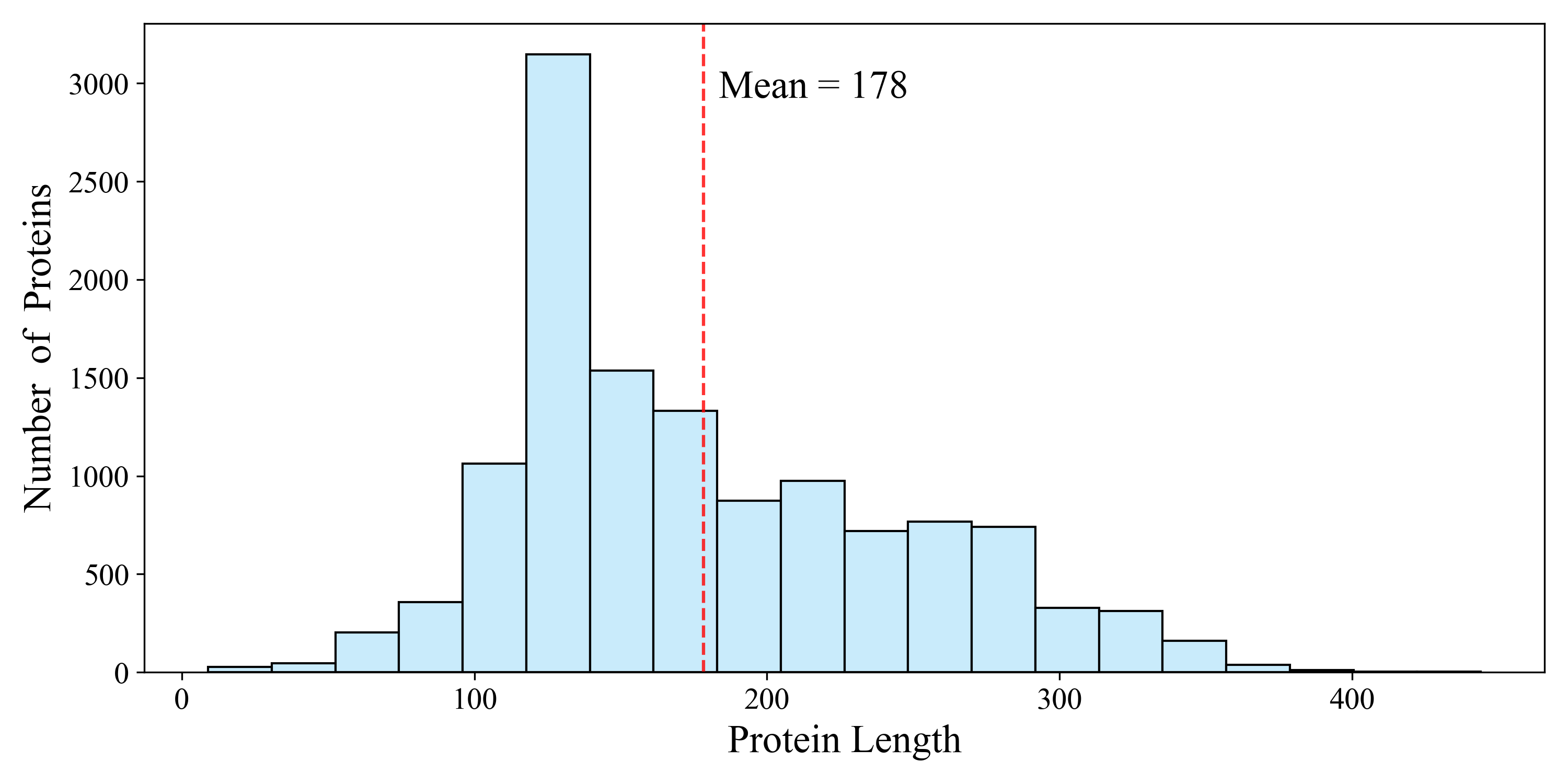}
        \includegraphics[width=1.0\textwidth]{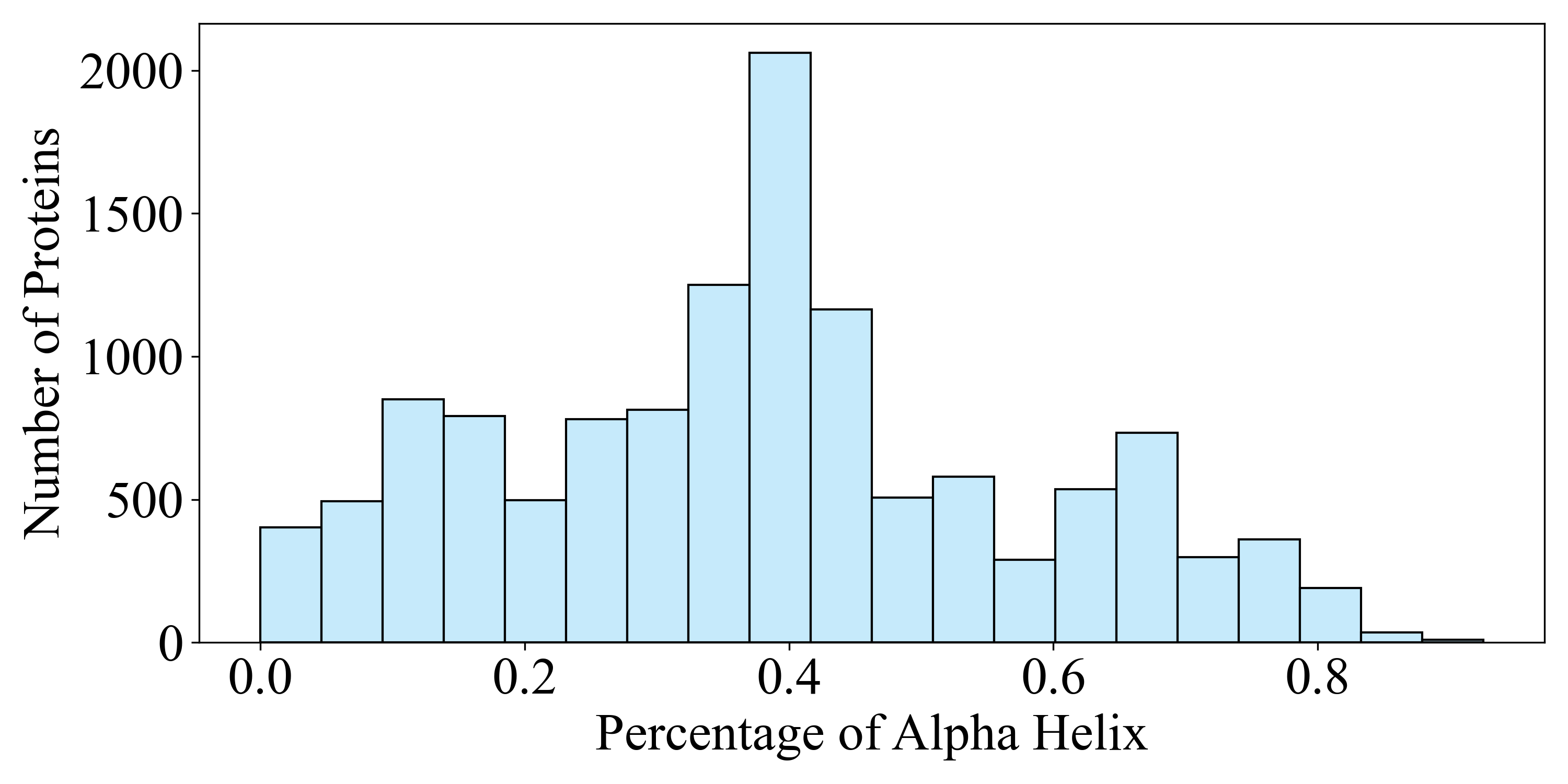}
       \end{minipage}
    \begin{minipage}[b]{0.43\textwidth}   
        \includegraphics[width=1.0\textwidth]{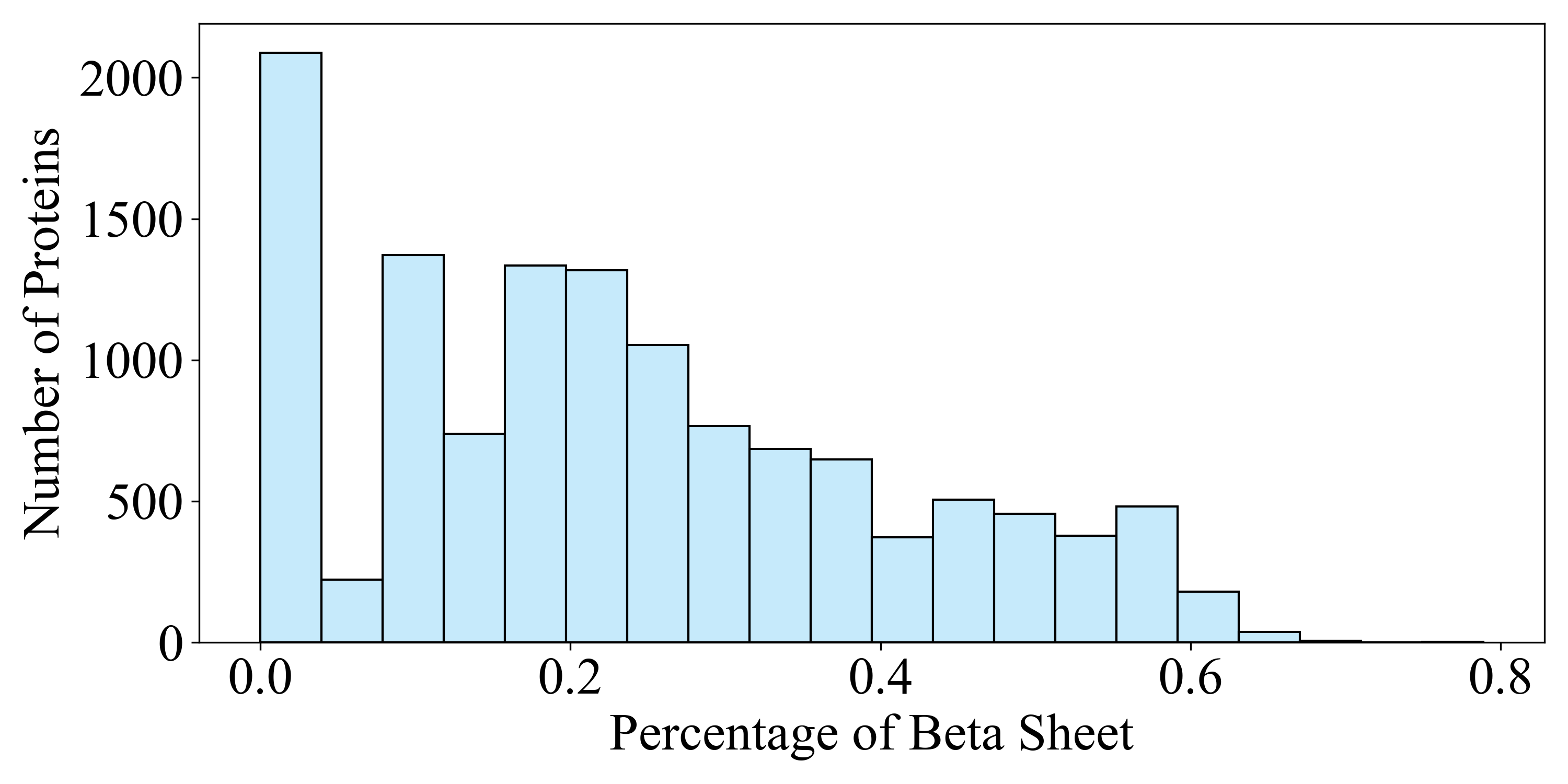}
        \includegraphics[width=1.0\textwidth]{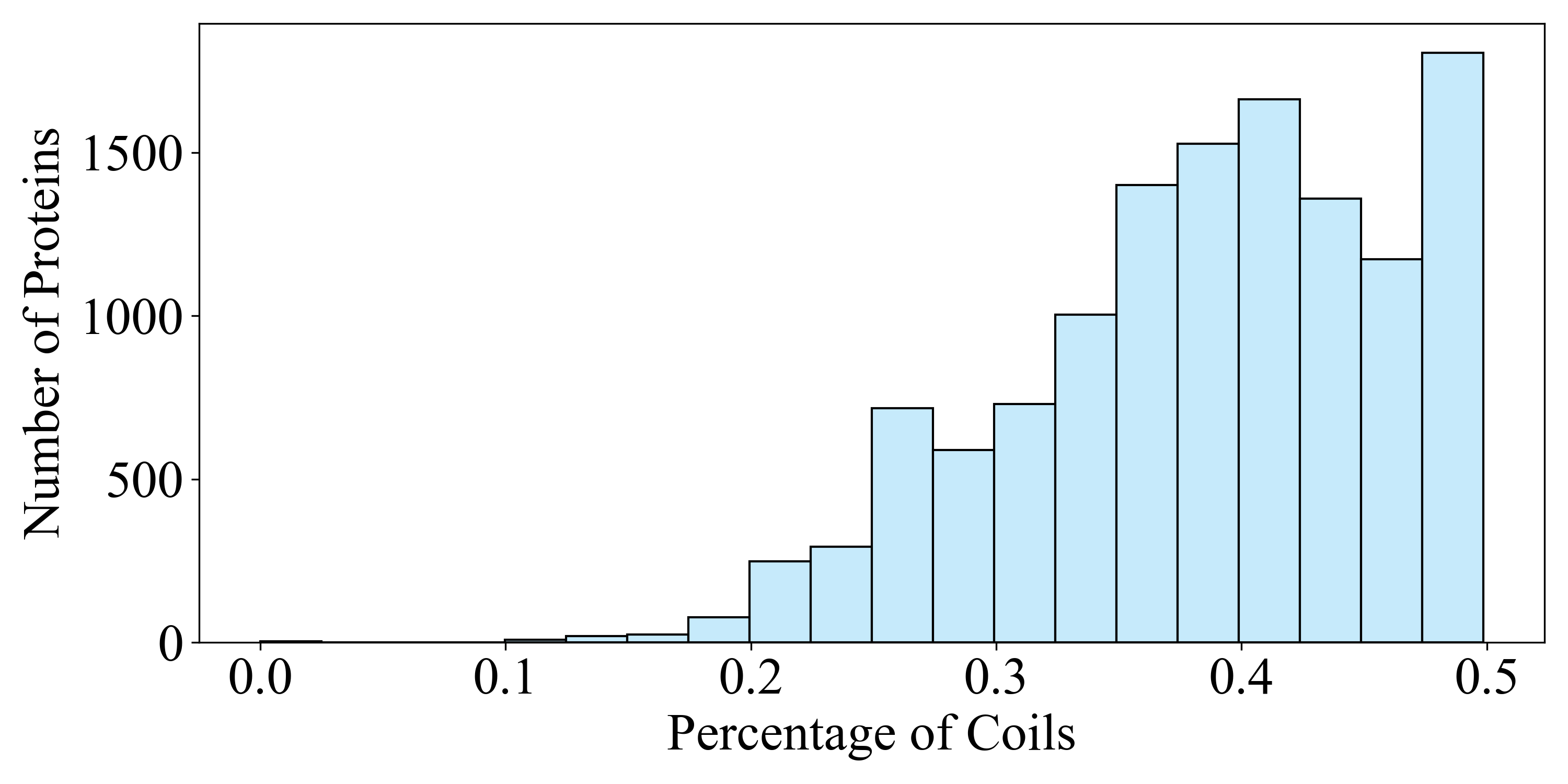}
        \end{minipage}   
    \caption{Overall statistics of the dynamic protein dataset on protein length and secondary structure.}
    \label{fig:statistics_structure}
\end{figure*}

\begin{table*}[h]
\centering
\begin{tabular}{c|c|c|c} 
\hline
    Name & Type & Description & Unit\\
\hline
    Protein ID & string & Identifier of Protein & -- \\
    
    Position & float array & Trajectory Coordinates & \AA \\
    
    Velocity & float array & Trajectory Velocities & \AA/ps \\
    
    Force & float array & Trajectory Forces & $\text{kcal/mol} \cdot \text{\AA}$ \\
    
    Potential Energy & float & System Potential Energy & kJ/mole \\
    
    Kinetic Energy & float & System Kinetic Energy & kJ/mole \\  
    
    Total Energy & float  & System Total Energy & kJ/mole \\ 
    
    Temperature & float & System Temperature & K \\
    
    Box Volume & float  & System Volume Forces & $\text{nm}^3$ \\ 
    
    Density & float & System Density & g/mL \\

    Final Status & xml & Status for Prolongation& -- \\
    
\hline
\end{tabular}
\caption{Attributes of proposed dataset.}
\label{tab:simulationoutput_sup}
\end{table*}

\paragraph{Statistics of Protein Completion Methods.} In Figure~\ref{fig:completion_percent}, we present the adopted methods to complete the missing residues of proteins, and the percentage of each method used. Specifically, around 33\% proteins are complete and don't require completion. Around 23\% proteins are completed by MODELLER, and around 43\% are completed by AlphaFold 2.

\paragraph{Attributes in Dynamic Protein Dataset.}
Table~\ref{tab:simulationoutput_sup} provides a detailed overview of the data attributes associated with each protein. This structured format supports subsequent analyses and interpretations of the intrinsic dynamic behaviors and properties of the proteins within the dataset.

\begin{figure}[h]
\centering
\includegraphics[width=1.0\linewidth]{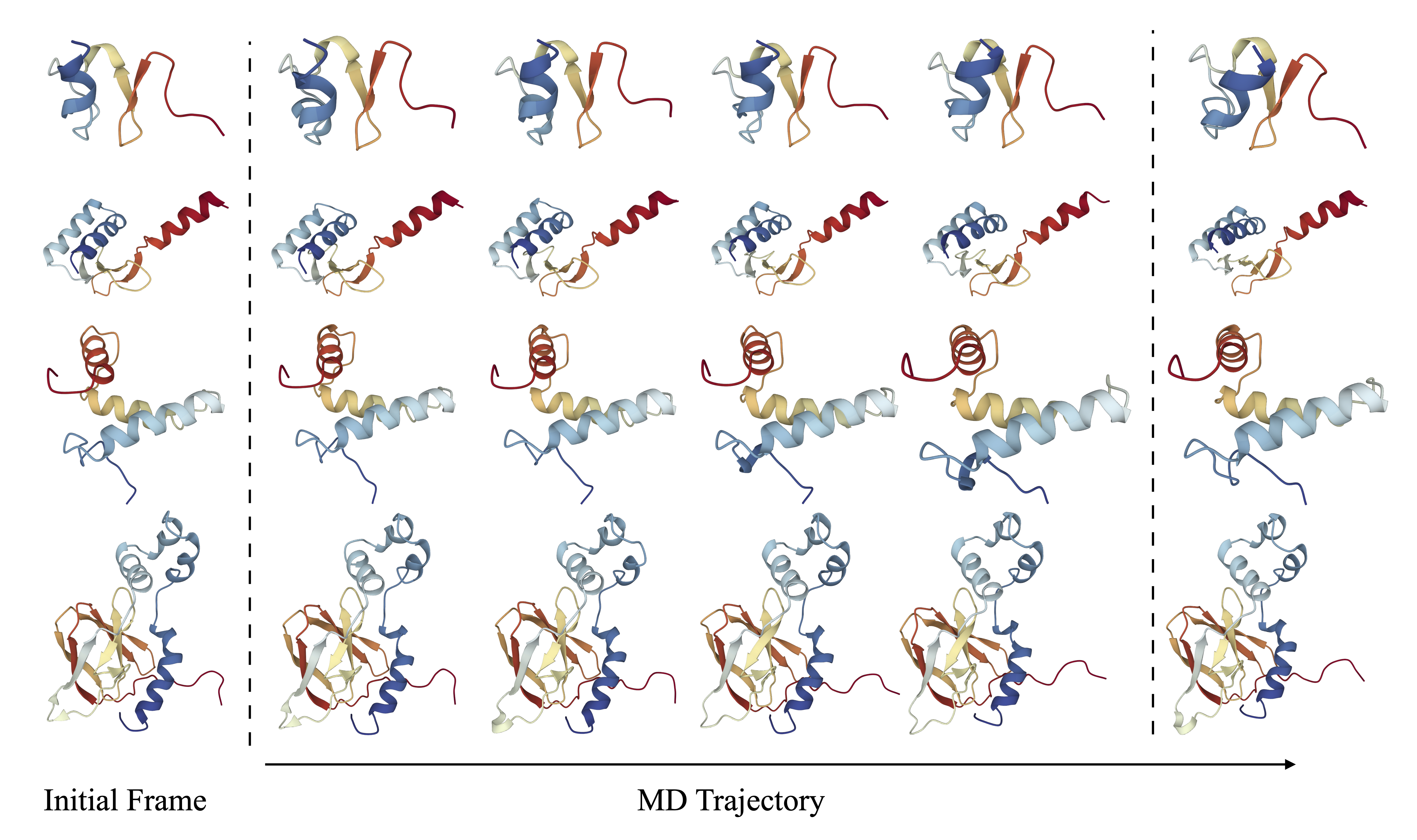}
\caption{Visualization of our protein trajectories, stored with higher temporal resolution, offers a more detailed depiction of the protein's trajectories.} 
\label{fig:high_resolution_sup}
\end{figure}

\begin{figure*}[t]
    \centering
    \subfigure[Force]{
        \begin{minipage}[b]{0.23\textwidth}
        \centering
        \includegraphics[width=1.0\textwidth]{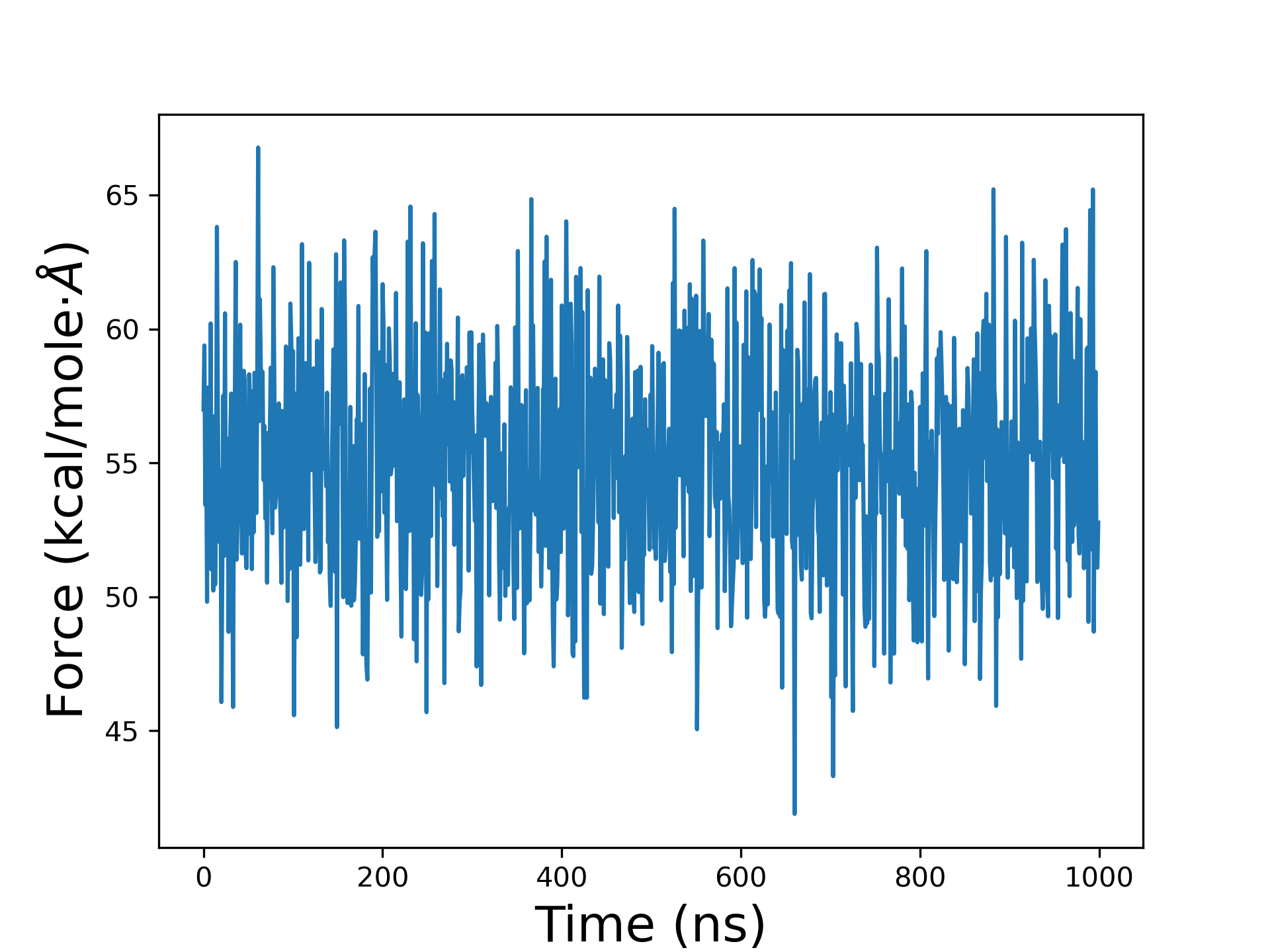}
        \end{minipage}
    }
    \subfigure[Velocity]{
        \begin{minipage}[b]{0.23\textwidth}
        \centering
        \includegraphics[width=1.0\textwidth]{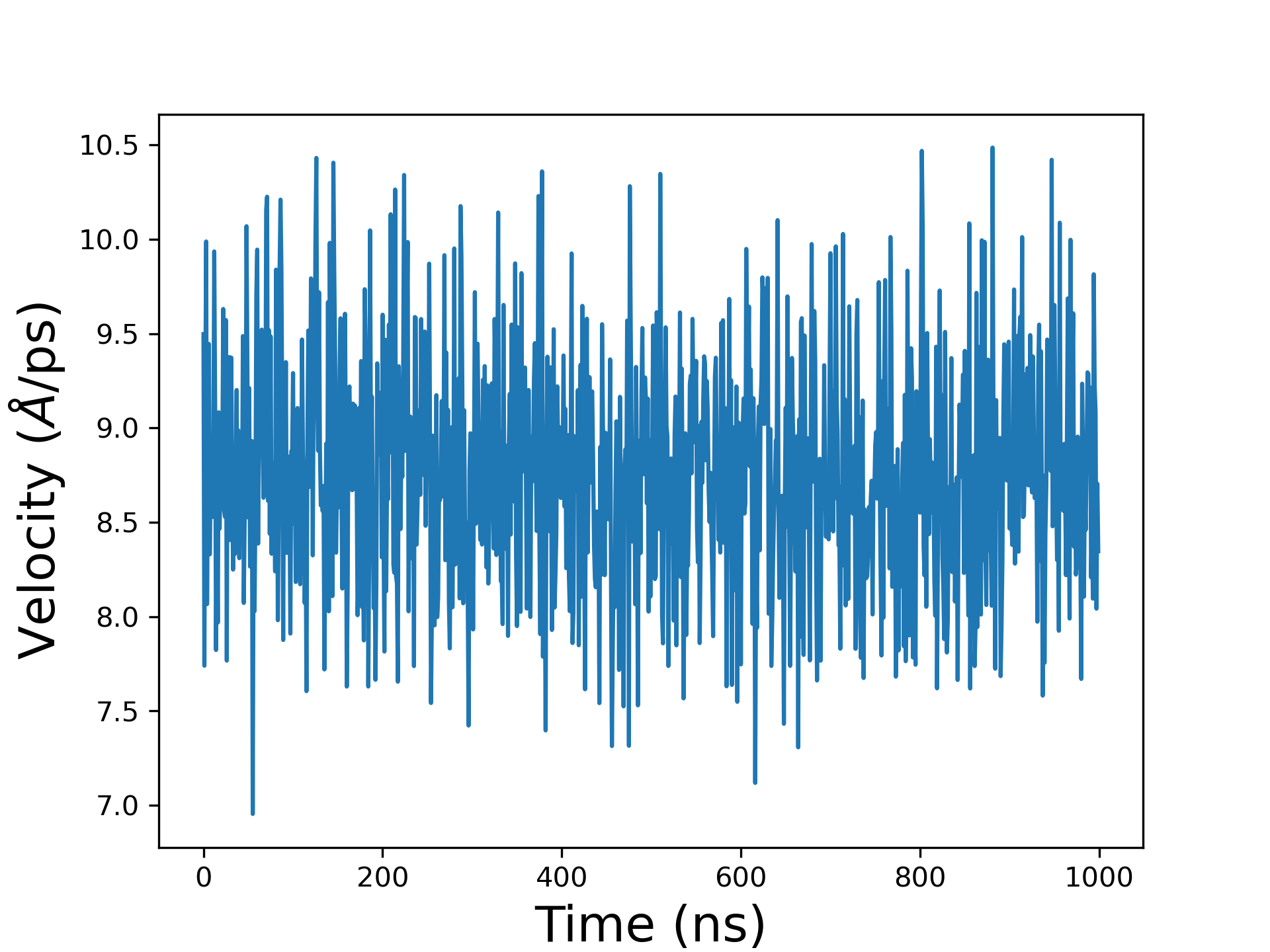}
        \end{minipage}
    }
    \subfigure[Temperature]{
        \begin{minipage}[b]{0.23\textwidth}
        \centering
        \includegraphics[width=1.0\textwidth]{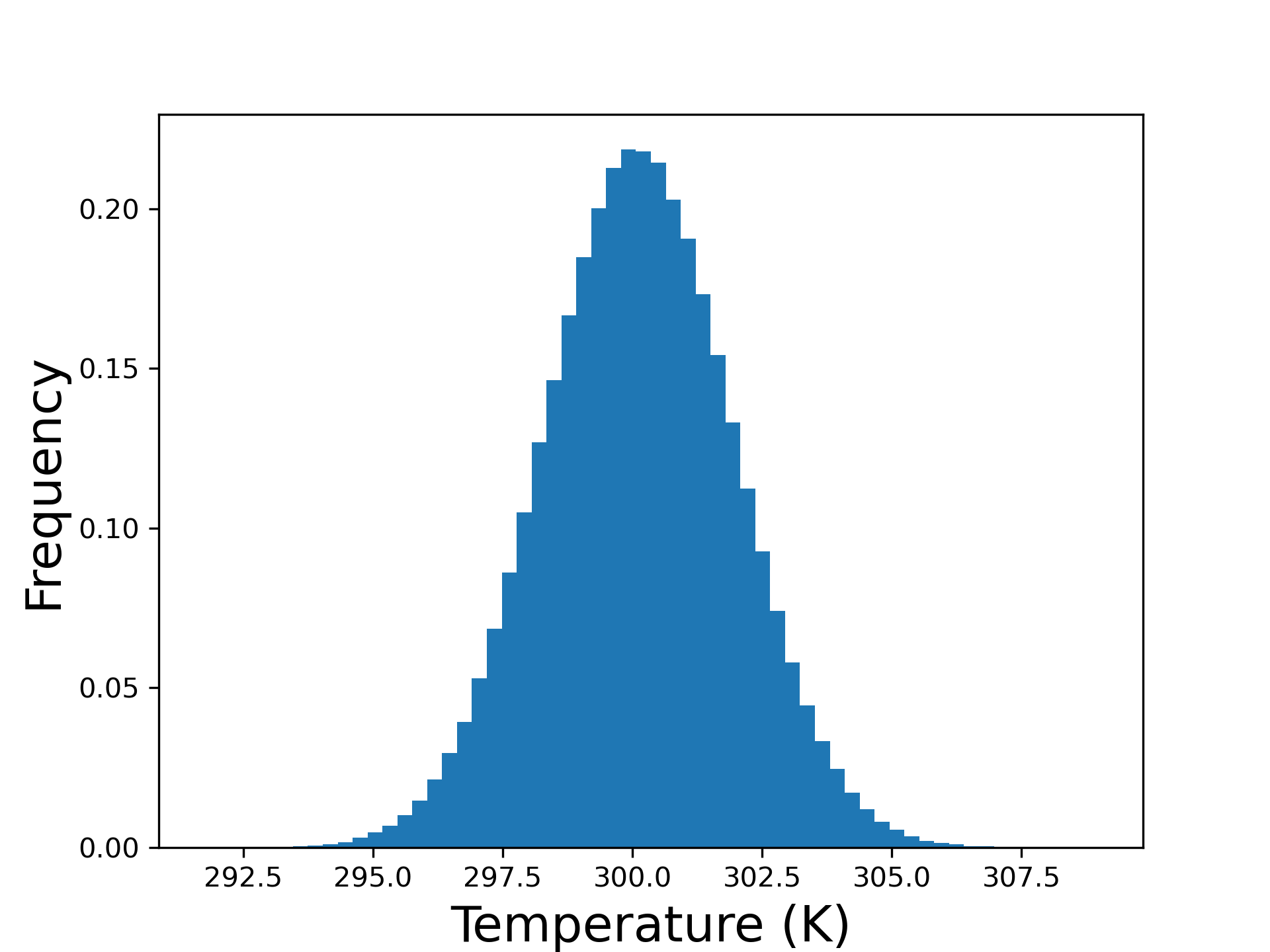}
        \end{minipage}
    }
    \subfigure[Potential Energy]{
        \begin{minipage}[b]{0.23\textwidth}
        \centering
        \includegraphics[width=1.0\textwidth]{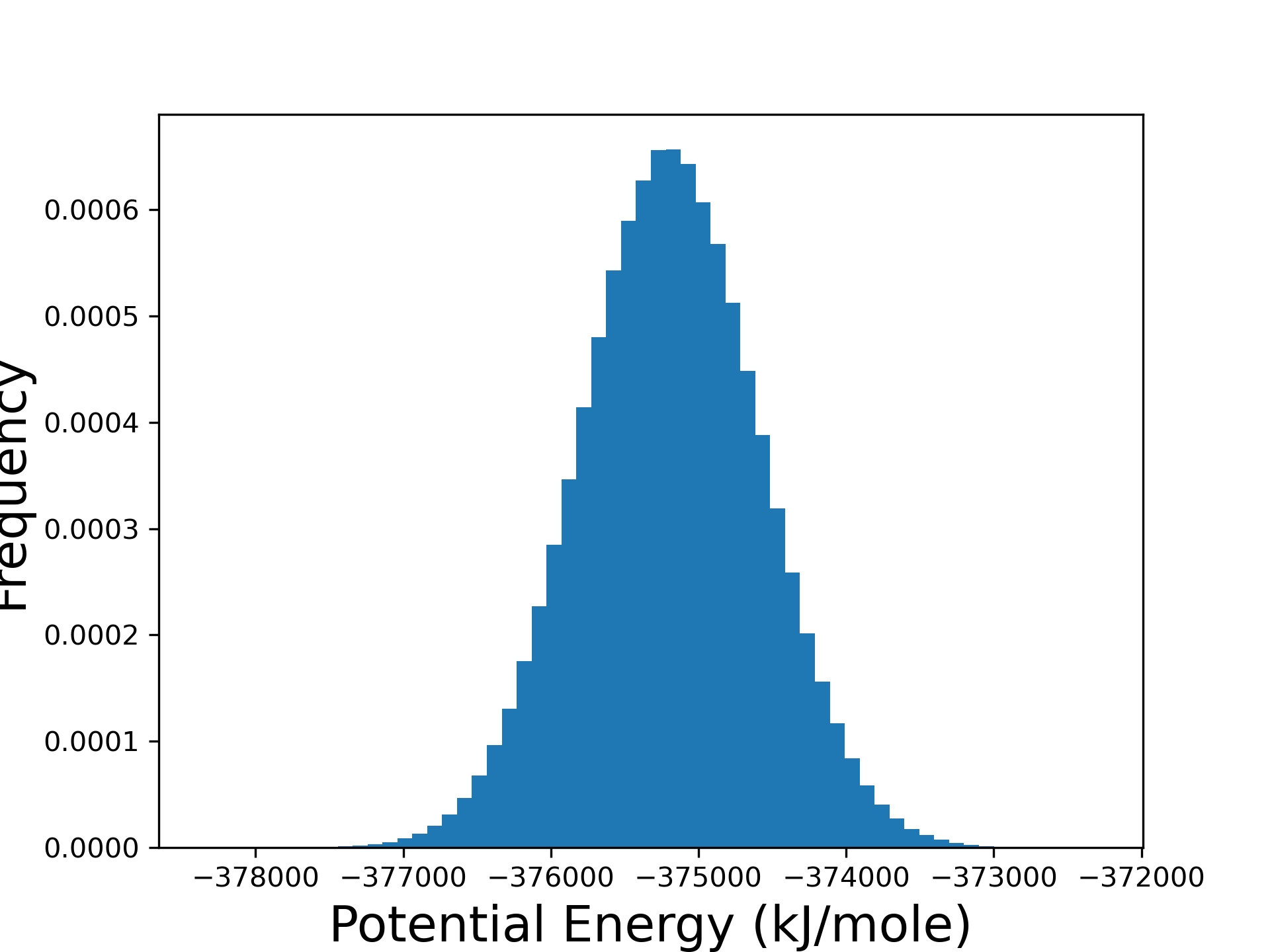}
        \end{minipage}
    }
    \caption{Statistics of the protein 3TVJ\_I on MD forces, velocities, temperature and potential energy from force field.}
    \label{fig:statistics_physics}
\end{figure*}
\paragraph{Statistics of Protein Structures.} We present statistics regarding protein length and secondary structures in Figure~\ref{fig:statistics_structure}, which illustrates the structural diversity present within the dataset. The protein lengths range from 9 to 444 residues. The percentage of alpha helices varies from 0.00 to 0.92. The percentage of beta sheets varies from 0.00 to 0.78. The percentage of coils varies from 0.00 to 0.49.
\\
\paragraph{Statistics of Physical Properties.} In Figure~\ref{fig:statistics_physics}, we present statistics about the physical properties of 3TVJ\_I, including force, velocity, temperature, and potential energy. In Figure~\ref{fig:statistics_physics} (a) and (b), we plot the curves of the norm of force and velocity averaged over $\mathtt{C}_{\alpha}$ atoms with respect to the simulation time. In Figure~\ref{fig:statistics_physics} (c) and (d), we plot the histogram of the temperature and potential energy of the system during the simulation. 
ß

\begin{figure*}[t]
\centering
\includegraphics[width=1.0\linewidth]{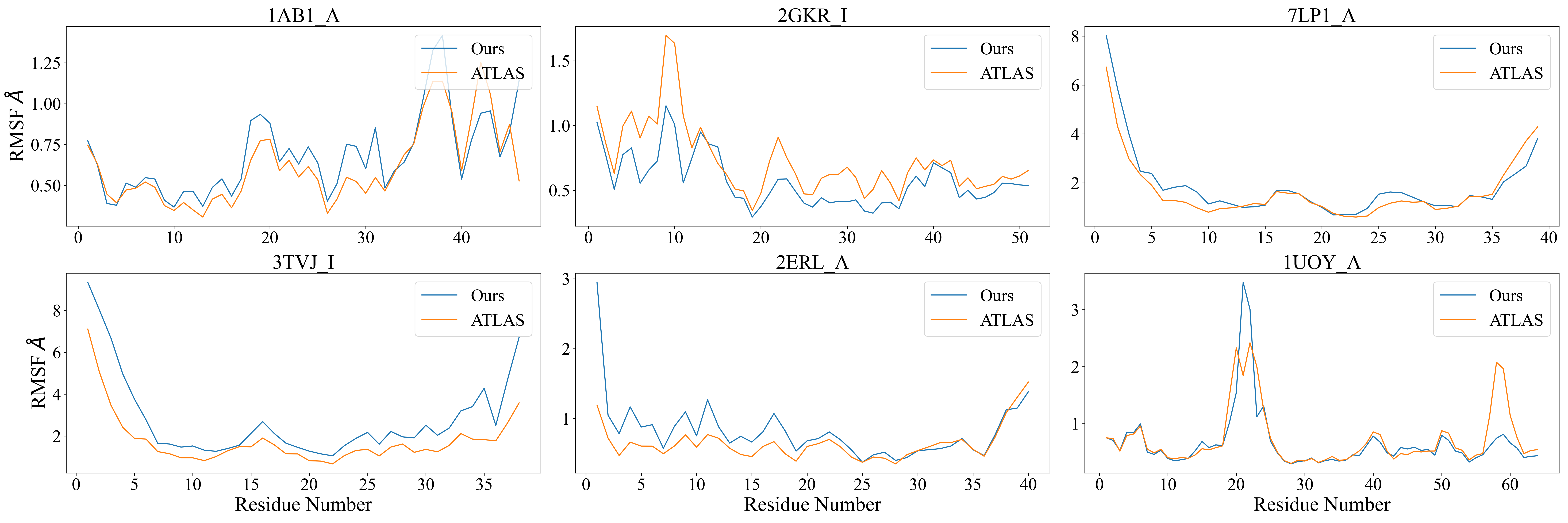}
\caption{RMSF comparison between the proposed dataset and ATLAS reveals similar residue fluctuations, effectively capturing the intrinsic dynamics of proteins.}
\label{fig:rmsf_comparison_sup}
\end{figure*}

\begin{figure*}[t]
\includegraphics[width=1.0\linewidth]{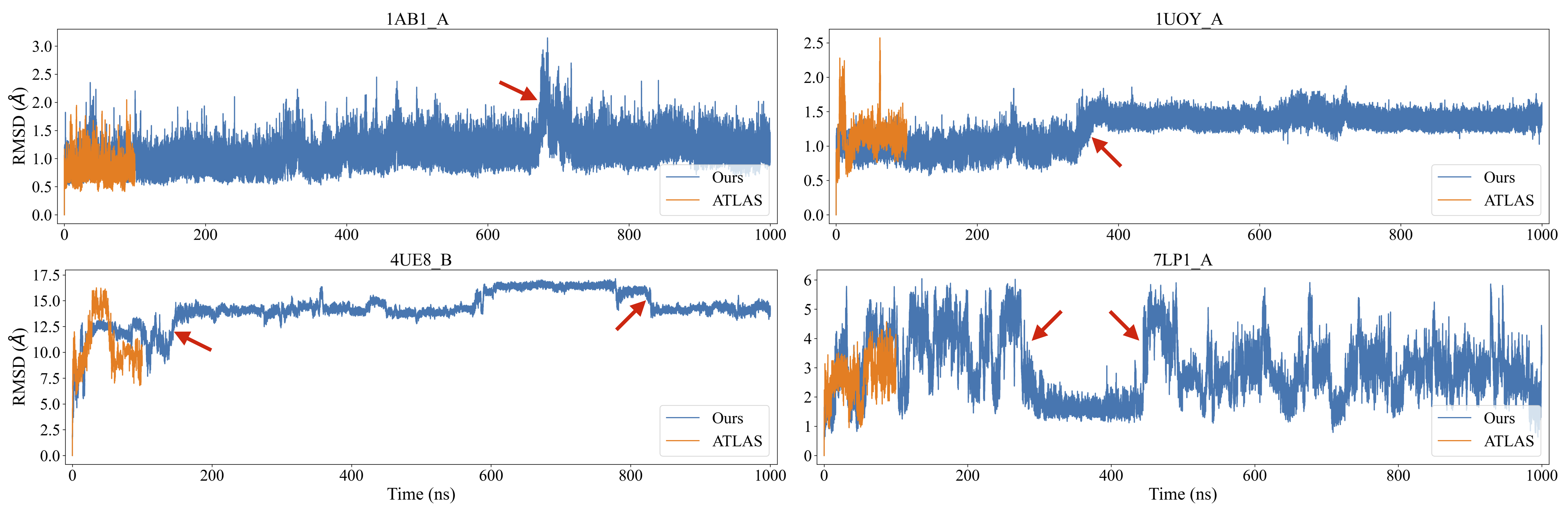}
\caption{RMSD plots for our dataset and ATLAS. Longer simulation time can potentially capture more protein conformational changes, which are indicated by the red arrows.}
\label{fig:rmsd_sup}
\end{figure*}

\begin{figure*}[t]
\centering
\includegraphics[width=0.85\textwidth]{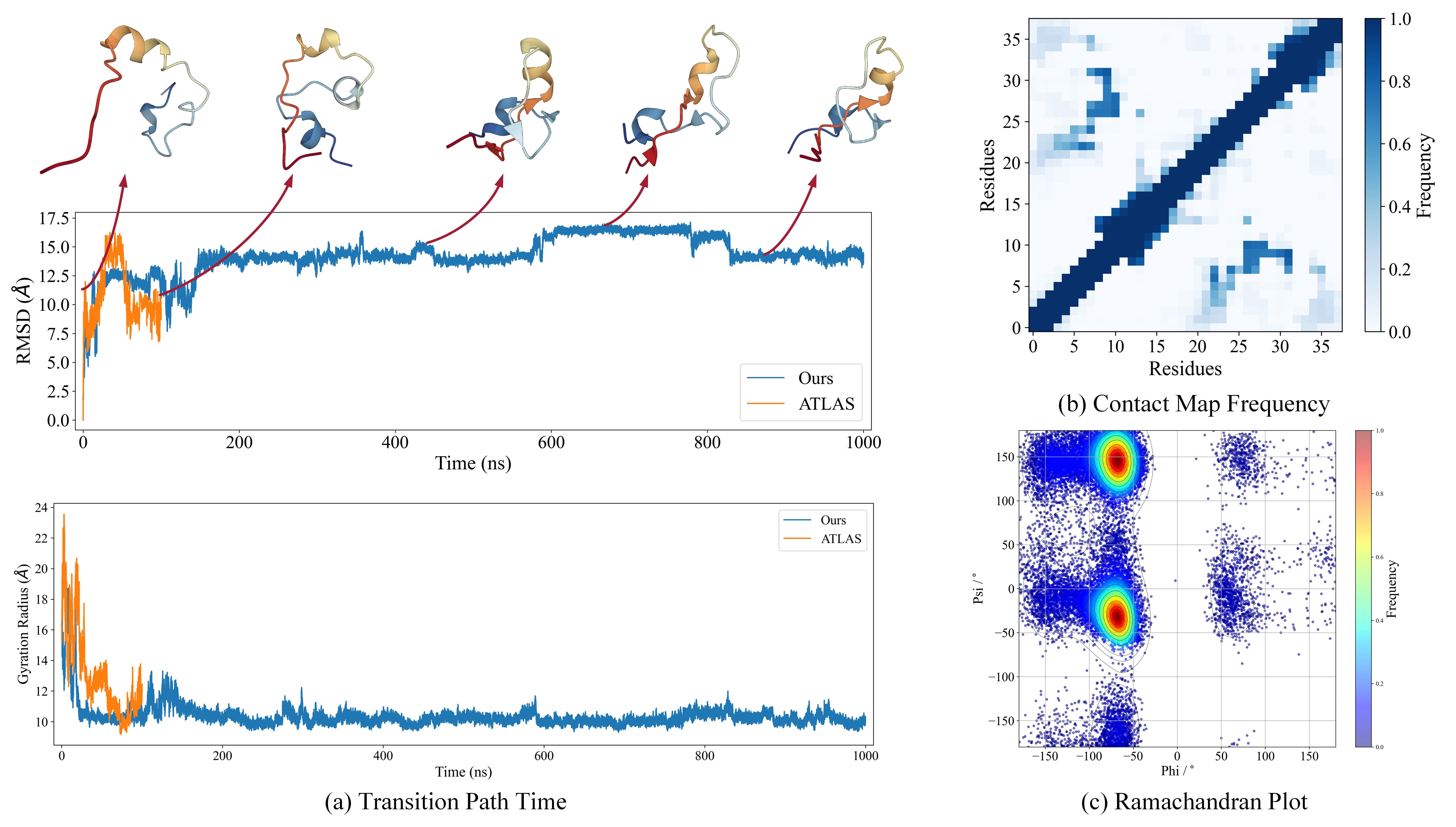}
\caption{The conformational evolution and statistics of protein 4UE8\_B from proposed dataset. a) The regions with the most significant changes in the RMSD (Root Mean Square Deviation) and radius of gyration curves over time correspond to potential conformational changes, as depicted in the upper part of the figure. b) The contact map frequency illustrates the changes in interactions between residues within the protein. c) The Ramachandran plot provides insight into the dihedral angles of the protein backbone, indicating the structural validity of the protein conformation.}
\label{fig:conf_evo1}
\end{figure*}

\begin{figure*}[!t]
\centering
\includegraphics[width=0.85\textwidth]{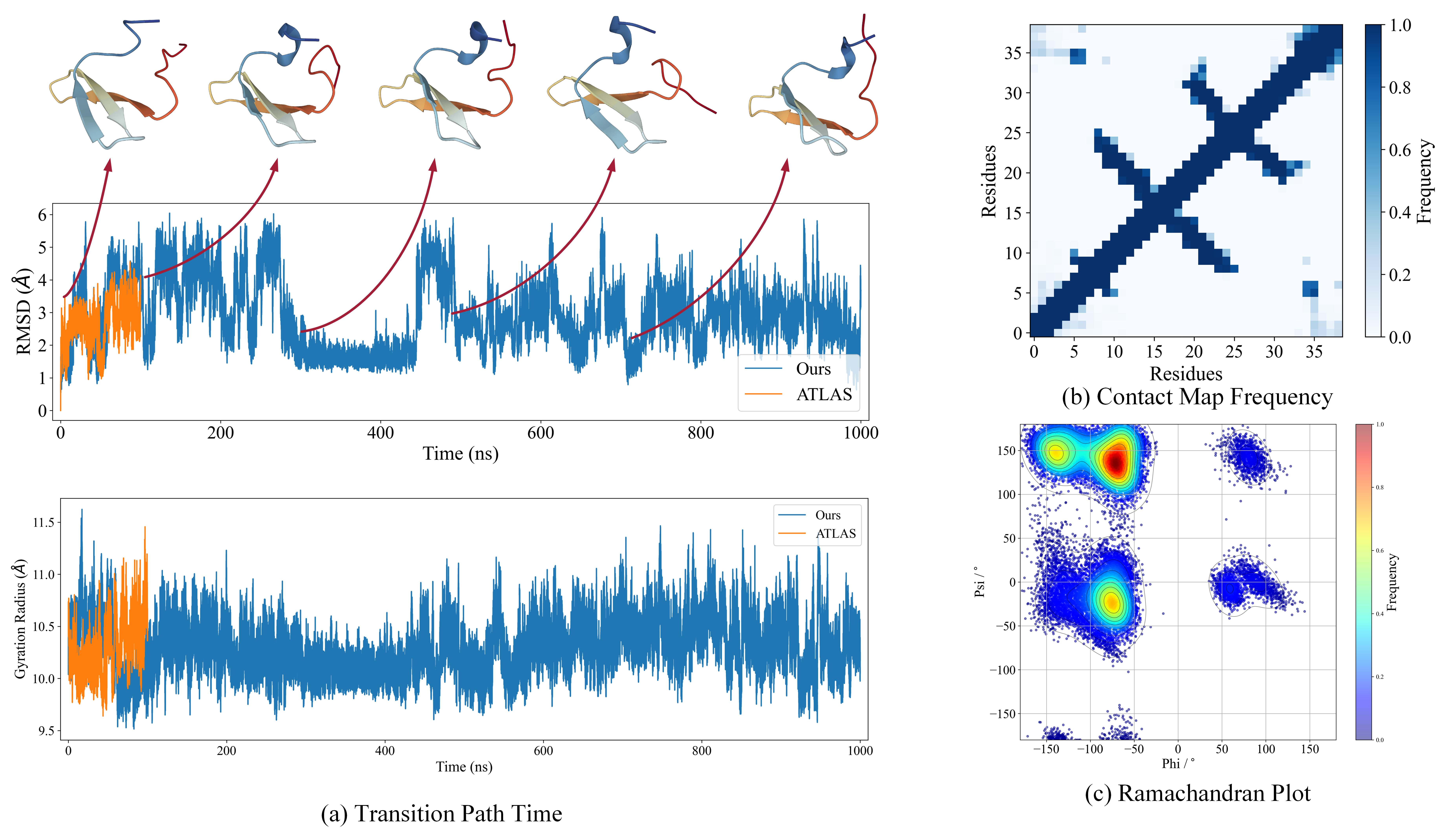}
\caption{The conformational evolution and statistics of protein 7LP1\_B from proposed dataset. a) The regions with the most significant changes in the RMSD (Root Mean Square Deviation) and radius of gyration curves over time correspond to potential conformational changes, as depicted in the upper part of the figure. b) The contact map frequency illustrates the changes in interactions between residues within the protein. c) The Ramachandran plot provides insight into the dihedral angles of the protein backbone, indicating the structural validity of the protein conformation.}
\label{fig:conf_evo2}
\end{figure*}

\paragraph{Analysis of Dynamic Behaviors.} We present additional protein examples to compare with ATLAS using RMSF and RMSD, as shown in Figure~\ref{fig:rmsf_comparison_sup} and Figure~\ref{fig:rmsd_sup}, respectively. The fluctuations of the protein residues in our dataset is consistent with the findings observed in ATLAS. However, the extended simulation time provides a more comprehensive set of conformational changes, as further illustrated by the visualization of the dynamic characteristics of 4UE8\_B and 7PL1\_B in Figure~\ref{fig:conf_evo1} and Figure~\ref{fig:conf_evo2}, respectively. We also present more examples of trajectories in Figure~\ref{fig:high_resolution_sup}. Higher temporal resolution can capture the subtle conformational changes of the protein, offering a more detailed depiction of the protein's trajectories.


\end{document}